\documentclass[aps,prd,superscriptaddress,showkeys,showpacs,twocolumn,longbibliography,nofootinbib,floatfix]{revtex4-1}
\usepackage[utf8]{inputenc}
\usepackage{amsmath}
\usepackage{amsfonts}
\usepackage{amsthm}
\usepackage{amssymb}
\usepackage{float}
\usepackage{braket}
\usepackage{graphicx}
\usepackage{xcolor}
\usepackage{url}
\usepackage[colorlinks=true, pdfstartview=FitV, linkcolor=red, citecolor=blue, urlcolor=blue]{hyperref}
\usepackage{slashed}
\usepackage[normalem]{ulem}

\graphicspath{{./figures/}}

\newcommand{\be}{\begin{equation}}
\newcommand{\ee}{\end{equation}}
\newcommand{\beq}{\begin{equation}}
\newcommand{\eeq}{\end{equation}}
\newcommand{\bea}{\begin{eqnarray}}
\newcommand{\eea}{\end{eqnarray}}

\newcommand{\Tr}{\mathrm{Tr}}

\newcommand{\Ldyn}[1]{L_{\mathrm{dyn},#1}}
\newcommand{\Lext}[1]{L_{\mathrm{ext},#1}}

\begin{document}
\title{Quantum simulation of entanglement and hadronization in jet production:\\ lessons from the massive Schwinger model}

\author{Adrien Florio}
\email[]{aflorio@bnl.gov}
\affiliation{Department of Physics, Brookhaven National Laboratory, Upton, New York 11973-5000, USA}
\affiliation{Co-design Center for Quantum Advantage}

\author{David Frenklakh}
\email[]{david.frenklakh@stonybrook.edu}
\affiliation{Center for Nuclear Theory, Department of Physics and Astronomy, Stony Brook University, Stony Brook, New York 11794-3800, USA}

\author{Kazuki Ikeda}
\email[]{kazuki.ikeda@stonybrook.edu}
\affiliation{Co-design Center for Quantum Advantage}
\affiliation{Center for Nuclear Theory, Department of Physics and Astronomy, Stony Brook University, Stony Brook, New York 11794-3800, USA}

\author{\mbox{Dmitri Kharzeev}}
\email[]{dmitri.kharzeev@stonybrook.edu}
\affiliation{Department of Physics, Brookhaven National Laboratory, Upton, New York 11973-5000, USA}
\affiliation{Co-design Center for Quantum Advantage}
\affiliation{Center for Nuclear Theory, Department of Physics and Astronomy, Stony Brook University, Stony Brook, New York 11794-3800, USA}

\author{Vladimir Korepin}
\email[]{vladimir.korepin@stonybrook.edu}
\affiliation{Co-design Center for Quantum Advantage}
\affiliation{C.N. Yang Institute for Theoretical Physics, Stony Brook University, Stony Brook, New York, 11794-3840, USA}

\author{Shuzhe Shi}
\email[]{shuzhe-shi@tsinghua.edu.cn}
\affiliation{Center for Nuclear Theory, Department of Physics and Astronomy, Stony Brook University, Stony Brook, New York 11794-3800, USA}
\affiliation{Department of Physics, Tsinghua University, Beijing 100084, China}

\author{Kwangmin Yu}
\email[]{kyu@bnl.gov}
\affiliation{Computational Science Initiative, Brookhaven National Laboratory, Upton, New York 11973-5000, USA}

\bibliographystyle{unsrt}

\begin{abstract}
The possible link between entanglement and thermalization, and the dynamics of hadronization are addressed by studying the real-time response of the massive Schwinger model coupled to external sources. This setup mimics the production and fragmentation of quark jets, as the Schwinger model and QCD share the properties of confinement and chiral symmetry breaking. By using quantum simulations on classical hardware, we study the entanglement between the produced jets, and observe the growth of the corresponding entanglement entropy in time. This growth arises from the increased number of contributing eigenstates of the reduced density matrix with sufficiently large and close eigenvalues. We also investigate the physical nature of these eigenstates, and find that at early times they correspond to fermionic Fock states. We then observe 
the transition from these fermionic Fock states to meson-like bound states as a function of time. In other words, we observe how hadronization develops in real time. At late times, the local observables at mid-rapidity (such as the fermion density and the electric field) approach approximately constant values, suggesting the onset of equilibrium and approach to thermalization. 
\end{abstract}

\maketitle

\section{Introduction}

The role of quantum entanglement in high energy hadronic interactions has recently excited a lot of interest~\cite{Kharzeev:2017qzs,Tu:2019ouv,Hentschinski:2023izh,H1:2020zpd,Kharzeev:2021nzh,Armesto:2019mna,Kharzeev:2021yyf,Dvali:2021ooc,Zhang:2021hra,Liu:2022ohy,Liu:2022hto,Dumitru:2023qee}. In particular, it has been found that QCD evolution produces at large rapidity separation $y$ (or, equivalently, at small Bjorken $x=e^{-y}$) a maximally entangled state (MES) \cite{Kharzeev:2017qzs}. This MES contains a large number of microstates with $n$ partons that possess approximately equal probabilities $p_n = p_n(y)$, giving rise to a maximal value of the entanglement entropy (EE)
\begin{equation}\label{EE}
    S_E(y) = - \sum_n p_n \ln p_n .
\end{equation}
The EE naturally emerges at high energies because the phase of the hadron wave function cannot be determined during the short Lorentz-contracted time of the interaction, so the measured density matrix corresponds to the density matrix of a pure hadron state traced over the unobserved phases \cite{Kharzeev:2021nzh}. 
\vskip0.3cm

One may further assume that the von Neumann entropy of the final hadron state $S_H$ is equal to the entanglement entropy~\eqref{EE}~\cite{Kharzeev:2017qzs,Tu:2019ouv,Hentschinski:2023izh}:
\begin{equation}\label{lphd}
    S_E = S_H .
\end{equation}
This can be viewed as a stronger form of the ``local parton-hadron duality" \cite{Azimov:1984np,Dokshitzer:1991eq} that emerges if one assumes that hadron multiplicity distribution mirrors the distribution in the number of partons measured at a given rapidity. Once this assumption is made, we get a relation between the parton structure function (dominated by gluons at small $x$), $x\, G(x) = {\bar n} = \sum_n n\ p_n$ and the entropy of hadronic final state that should hold for a MES~\cite{Kharzeev:2017qzs}:
\begin{equation}\label{MES}
    S_H = \ln[x\, G(x)]
\end{equation}
that can be directly tested in experiments \cite{Tu:2019ouv,H1:2020zpd,Hentschinski:2023izh}.
\vskip0.3cm

There is growing experimental evidence from deep inelastic scattering and high energy hadron interactions that the relation (\ref{MES}) indeed holds \cite{Tu:2019ouv,H1:2020zpd,Kharzeev:2021yyf,Hentschinski:2021aux}. While the validity of (\ref{MES}) imposed by the emergence of MES is intriguing, it raises several questions that currently remain unanswered:
\begin{enumerate}

    \item The (single) sum in (\ref{EE}) runs over the eigenstates of the reduced density matrix, or Schmidt states. It has been conjectured that these states correspond to parton states, but this conjecture was never proven, even for a weakly coupled theory. Things become even more complicated at strong coupling, when the notion of partons is ill-defined. What can be said from first principles about the nature of Schmidt states determining the EE?

    \item For a MES, the EE should be given by $S_E = \ln {\cal D}$, where ${\cal D}$ is the dimension of the Hilbert space in which the state is defined. In quantum field theory, this dimension is generally infinite so that the state can be maximally entangled only in a subspace of the full Hilbert space of the problem. What determines the dimension of this subspace? In more practical terms, what should be an upper limit in the sum~\eqref{EE}?

    \item Is it possible to establish the validity of relation~\eqref{lphd} in a solvable field-theoretic model with confinement?

    \item The theoretical illustrations of MES's emergence at high energies have been based on models without confinement. For example, the treatment in~\cite{Kharzeev:2017qzs} is based on BFKL dynamics that is conformally invariant. Will the MES emerge in a model with confinement? How will confinement affect the dimensionality of the Hilbert subspace mentioned above?

    \item There is strong experimental evidence in favor of an apparent thermalization in high energy collisions -- for example, hadron abundances can be viewed as thermal, with a universal effective temperature (see, e.g., \cite{Becattini:1997rv, Andronic:2008gu}). Is this apparent thermalization a consequence of the emergence of a MES? Is there a relation between an effective temperature and the confinement scale?
\end{enumerate}
\vskip0.3cm

To answer these questions, we need an explicit field theoretical model that shares with QCD the properties of confinement and chiral symmetry breaking, and in which the real-time evolution at high energies can be treated from first principles, either analytically or numerically. In this paper, we address this problem using a massive Schwinger model coupled to external sources. This setup models the production of high energy back-to-back jets in QCD, and has been studied before in \cite{Casher:1974vf,Loshaj:2011jx, Kharzeev:2012re, Kharzeev:2013wra, Florio:2023dke}. The massive Schwinger model cannot be solved analytically, 
but has been successfully studied using quantum simulations (see \cite{Bauer:2022hpo,Bauer:2023qgm} for recent reviews) on quantum \cite{Klco:2018kyo, Farrell:2023fgd,Farrell:2024fit} and classical \cite{Zache:2018cqq, Rigobello:2021fxw, deJong:2021wsd, Belyansky:2023rgh, Florio:2023dke, PhysRevD.108.L091501, Barata:2023jgd, PhysRevD.108.074001, Lee:2023urk} hardware. 

In this work we will use quantum simulations on classical hardware\footnote{Our real-time simulations require the quantum circuit depth that exceeds the capabilities of the current quantum computers. However, the  quantum simulations on quantum computers using our approach should become possible in the near future, see for instance \cite{Farrell:2023fgd,Farrell:2024fit}
}. In particular, we will complete exact diagonalization simulations on small lattices with tensor network simulations on larger lattices. 

\section{The model}
We study the massive Schwinger model with an external source. This external source consists of a fermion and antifermion going back to back along the lightcone, starting at the center of the lattice. As these external charges separate, the electric field between them is screened by dynamical fermion-antifermion pair production. This process resembles jet fragmentation in QCD and this version of Schwinger model modified by external sources was in fact suggested quite some time ago \cite{Casher:1974vf}. Using open-boundary conditions, the remaining gauge freedom allows to set the gauge potential to zero and express the electric field only in terms of fermionic operators by solving Gauss' law, see for instance \cite{Ikeda:2020agk} for explicit expressions. Using staggered fermions, our time dependent Hamiltonian reads (see \cite{Florio:2023dke} for more details)

\begin{align}
H_S^L &= -\frac{i}{2a}\sum_{n=1}^{N-1}
\big(\chi^\dag_{n}\chi_{n+1}-\chi^\dag_{n+1}\chi_{n}\big)+ m\sum_{n=1}^{N} (-1)^n \chi^\dag_n\chi_n 
\nonumber\\
&+\frac{ag^2}{2}\sum_{n=1}^{N-1}\left(\Ldyn{n}+\Lext{n}(t)\right)^2\ , \label{eq:Hamiltonian}
\end{align}
where $n$ labels lattice sites, going from 1 to the length of the lattice $N$; $\chi^\dagger_n$ are fermionic creation operators, $a$ is the lattice spacing, $m$ and $g$ are the fermion mass and electric charge, respectively. In the last term we use
\begin{align}
  \Ldyn{n} &= \sum_{i=1}^n q_i\,, \label{eq:E_dyn}\\
  \Lext{n}(t) &= -\theta\left(\frac{t-t_0}{a} - \left|n-\frac{N}{2}\right|\right) \ , \label{eq:E_ext}
\end{align}
where the first equation defines the dynamical lattice electric field operator in terms of the local charge operators (which is possible because of the one spatial dimension and the Gauss law) $q_i = n_i + \frac{(-1)^i-1}{2}$ with $n_i=\chi^\dagger_i \chi_i$ the fermion number operator.  The second equation describes the electric field due to the external sources propagating back to back along the lightcone. 

To study jet propagation in the vacuum, we begin the simulation in the vacuum state $|\Psi_0\rangle$ of the massive Schwinger model without the external source, described by the Hamiltonian $H(t=0)$. The  subsequent time evolution of this state is determined by the full Hamiltonian including the external source: 
\beq 
|\Psi_t\rangle = \mathcal{T}e^{-i \int_0^t dt' H(t')}|\Psi_0\rangle, \label{eq:state}
\eeq
where $\mathcal{T}$ denotes time ordering. We perform the time evolution numerically using exact diagonalization for lattice size $N\leq 20$.  For lattices of size $80\leq N\leq 100$, we use tensor networks methods. In more detail,  the Density Matrix Renormalization Group (DMRG)  is used to find a Matrix Product State (MPS) approximation of the ground state, which is subsequently evolved using the Time-Dependent Varational Principle (TDVP). We use the open-source ITensors library  \cite{itensor, itensor-r0.3} as our primary software source. Exact diagonalization preserves the full information about the state (\ref{eq:state}) and allows to extract any quantity characterizing the quantum state. Tensor networks allow to simulate much larger systems and surprisingly long-time evolution. Note that time evolution continues until external charges reach the edges of the lattice at $t = \frac{aN}{2}$ and then the simulation stops.

\section{Entanglement spectrum}

\begin{figure}
    \centering
    \includegraphics[width=\linewidth]{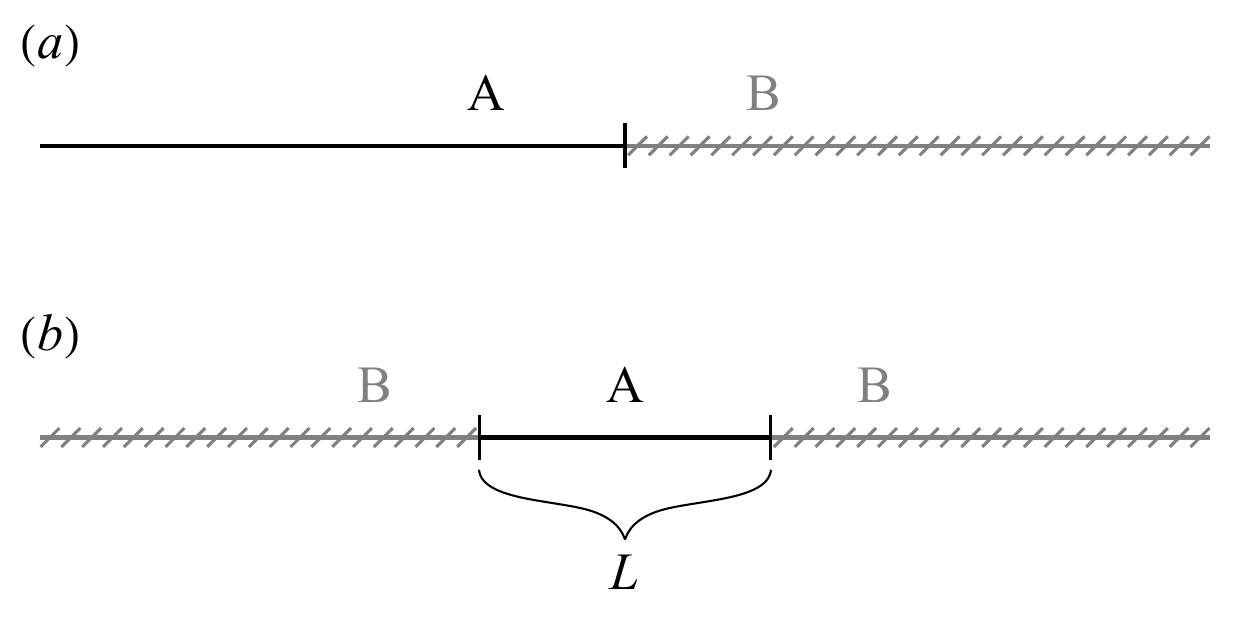}
    \caption{(a) A bipartition of the system for the analysis of entanglement between the two jets (left-right entanglement). (b) A bipartition of the system for the analysis of entanglement between the central region of length $L$ with the rest of the system.}
    \label{fig:entropy_AB}
\end{figure}

In order to study the entanglement between the two jets, we consider the partition of the full Hilbert space into two subspaces corresponding to the  left(L) and the right(R) halves of the lattice: $\mathcal{H} = \mathcal{H}_L\otimes\mathcal{H}_R$, see also Fig.\ref{fig:entropy_AB}, panel (a). Since one of the external charges propagates through $L$ and the other one through $R$, the entanglement between $\mathcal{H}_L$ and $\mathcal{H}_R$ corresponds to the entanglement between jets. It was previously studied in \cite{Florio:2023dke} how the entanglement entropy between $\mathcal{H}_L$ and $\mathcal{H}_R$ is generated. To have a better understanding of the entanglement structure we will now study the entanglement spectrum, also called the Schmidt spectrum. We first review how it is defined.

Starting with the quantum state $|\Psi_t\rangle$ given by Eq.~\eqref{eq:state}, one can define the density matrix $\rho(t) = |\Psi_t\rangle\langle\Psi_t|$ and perform the Schmidt decomposition:
\beq
\rho(t) = \sum_{i=1}^{2^{N/2}} \lambda_i(t) |\psi_i(t)\rangle\langle\psi_i(t)|, \label{eq:Schmidt_decomposition}
\eeq
where each of the states in the decomposition is a tensor product of a state in $\mathcal{H}_L$ and $\mathcal{H}_R$: 
\beq
|\psi_i(t)\rangle = |\psi_i^L(t)\rangle\otimes|\psi_i^R(t)\rangle.
\eeq
The states $\{|\psi_i^{L/R}(t)\rangle\}$ are called Schmidt vectors. They form orthonormal bases in the corresponding Hilbert subspaces. The coefficients of the decomposition (\ref{eq:Schmidt_decomposition}) $\lambda_i$ are called Schmidt coefficients. They can be interpreted as the eigenvalues of the reduced density matrix, e.g. for the left subsystem:
\beq
\rho_L(t) = \Tr_R \rho(t) = \sum_{i=1}^{2^{N/2}} \lambda_i(t) |\psi_i^L(t)\rangle\langle\psi_i^L(t)|,
\eeq
where $\Tr_R$ denotes a partial trace over the right subsystem. The states $\{|\psi_i^L(t)\rangle\}$ are thus the eigenvectors of the reduced density matrix.

Entanglement entropy between $\mathcal{H}_L$ and $\mathcal{H}_R$ can be expressed via Schmidt coefficients (which are in this context also called the entanglement spectrum~\cite{PhysRevB.81.064439,PhysRevD.108.L091501}):
\beq
S_{EE}(t) = -\Tr_L[\rho_L(t) \ln\rho_L(t)] = -\sum_{i=1}^{2^{N/2}} \lambda_i\ln\lambda_i.
\eeq

The entanglement spectrum provides a more detailed view on the structure of entanglement between the two subsystems. In particular, it allows for a straightforward computation of the R\'enyi entropy of any order $\alpha$:
\beq
S_\alpha(t) \equiv \frac{\ln\Tr_{L}(\rho_L(t)^\alpha)}{1-\alpha} = \frac{\ln\sum_{i=1}^{2^{N/2}}\lambda_i^\alpha}{1-\alpha}. \label{eq:Renyi}
\eeq

The system described by the Hamiltonian (\ref{eq:Hamiltonian}) possesses a conserved charge, namely the electric charge $Q = \sum_{n=1}^N q_n$. Since at $t=0$ the vacuum state is in the $Q=0$ sector, in the subsequent time evolution the state (\ref{eq:state}) stays in the same sector. Schmidt vectors have a certain charge as well, but it does not have to be zero, as long as the sum of $Q_L$ and $Q_R$ is zero. Here $Q_L$ is defined as:
\beq
\sum_{n=1}^{N/2} q_n |\psi_i^L\rangle \equiv Q_L |\psi_i^L\rangle,
\eeq
and similarly for $Q_R$.

\begin{figure}
\includegraphics[width=\linewidth]{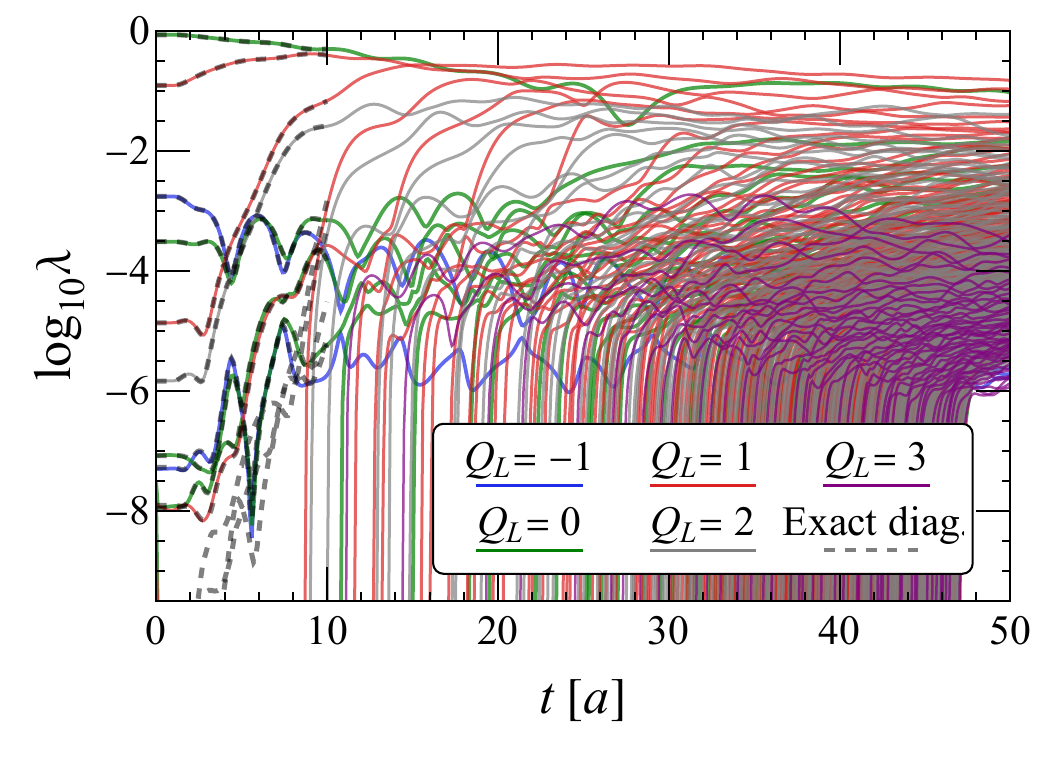}
\caption{Symmetry-resolved entanglement spectrum evolution for the lattice size $N=100$, $m=1/(4a), g=1/(2a)$. For comparison the spectrum obtained with exact diagonalization for $N=20$ at the same mass and coupling is shown as dashed curves.}
\label{fig:spectrum_N20_N100}
\end{figure}

Bipartite entanglement between two neighboring regions is readily computed from the MPS. Fig. \ref{fig:spectrum_N20_N100} displays the evolution of the Schmidt spectrum for $N=100$ lattice sites. At early times only very few Schmidt coefficients are large enough to contribute to the entanglement entropy. However, many more Schmidt coefficients emerge at later times and signal a step towards a partial maximal entanglement (in a maximally entangled state, all the Schmidt coefficients would be equal). The spectrum displayed is symmetry-resolved, i.e. we keep track of the electric charge of each eigenstate, $Q_L$. The maximal charge present in the spectrum is 3, corresponding to 3 quark-antiquark pairs cut by the entanglement surface. 

As a consistency check of the tensor network simulation we also compare this entanglement spectrum evolution with the one obtained using exact diagonalization on a smaller lattice, namely $N=20$. Since the evolution only continues until jets reach the boundary of the system, on the small lattice we only have access to relatively early times, but in the accessible domain the agreement is perfect, as one can see in Fig.~\ref{fig:spectrum_N20_N100}. This is a signature of short-ranged correlations as it demonstrates that entanglement between the two halves of the system is concentrated near the interface. As long as the time evolution in the near-interface region is the same, it does not matter what the system looks like far away from the interface. The discrepancy at the bottom of the spectrum is natural, as the tensor network approach inevitably contains a truncation in the Hilbert space, while the exact diagonalization approach captures all of the states. The consistency of the tensor network truncation is confirmed by the smallness of the omitted Schmidt coefficients. Moreover, once the corresponding coefficients grow enough, they are captured by the tensor network simulation.

\begin{figure}
\includegraphics[width=\linewidth]{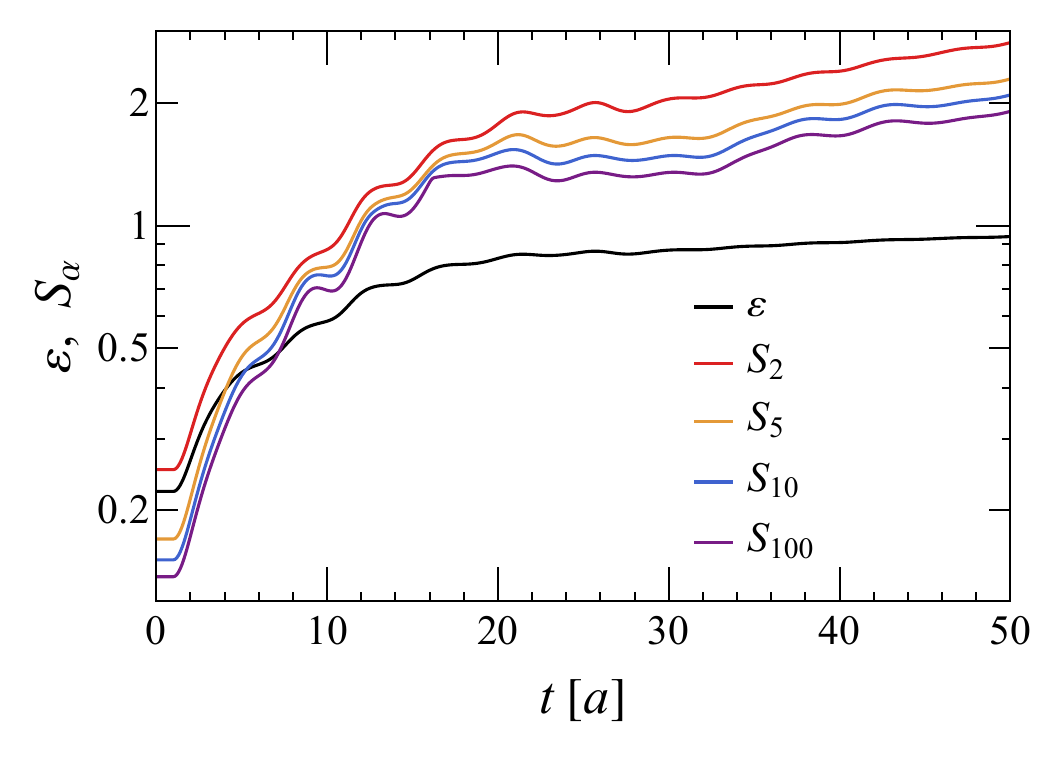}
\caption{Entangleness (black) and R\'enyi entropy with $\alpha=2$ (red), $5$ (gold), $10$ (blue), and $100$ (purple).}
\label{fig:Renyi}
\end{figure}

We then move on to gauge whether the density matrix ($\rho_L$) is closer to a pure state (PS) or a maximally entangled state (MES). 
We take two metrics to measure the level of entanglement: a) the R\'enyi entropy, defined in Eq.~\eqref{eq:Renyi} and b) the ``entangleness'':
\begin{align}
    \mathcal{E} \equiv \frac{1-\mathrm{tr}\rho_L^2}{1-2^{-N/2}}
    = \frac{1-\sum_{i=1}^{2^{N/2}} \lambda^2}{1-2^{-N/2}}\,.
\end{align}
It is worth noting that the entangleness is related to the variance ($\sigma^2$) of spectrum by $\mathcal{E} = 1- 2^{N/2}\,\sigma^2$.
For a PS, one of the $\lambda_i$'s equals to unity whereas all others vanish, $\mathrm{tr} \rho_L^2 = \mathrm{tr} \rho_L = 1$. Therefore, $S_{\alpha}[\mathrm{PS}] = 0$ and $\mathcal{E}[\mathrm{PS}] = 0$. 
Whereas for the MES, all $\lambda_i$ are equal to $2^{-N/2}$, $S_{\alpha}[\mathrm{MES}] = \frac{N \ln 2}{2}$ is independent of $\alpha$, and $\mathcal{E}[\mathrm{MES}] = 1$. 

In Fig.~\ref{fig:Renyi} we present our results for the entangleness and R\'enyi entropy. One can see that the entangleness  rapidly grows with time, and approaches saturation at a high value of $\simeq 0.9$ indicating the onset of a maximally entangled state. The left-right R\'enyi entropies $S_\alpha$ also rapidly increase with time, with a weak dependence on $\alpha$, again consistent with an approach to a MES. However, the absolute values of R\'enyi entropies are much smaller than $S_\alpha = \ln D$ expected for a MES in the entire Hilbert space of the system that has dimension $D=2^{N/2}$. This indicates that the MES is created in a particular subspace of the entire Hilbert space of the system.

\begin{figure}
\includegraphics[width=\linewidth]{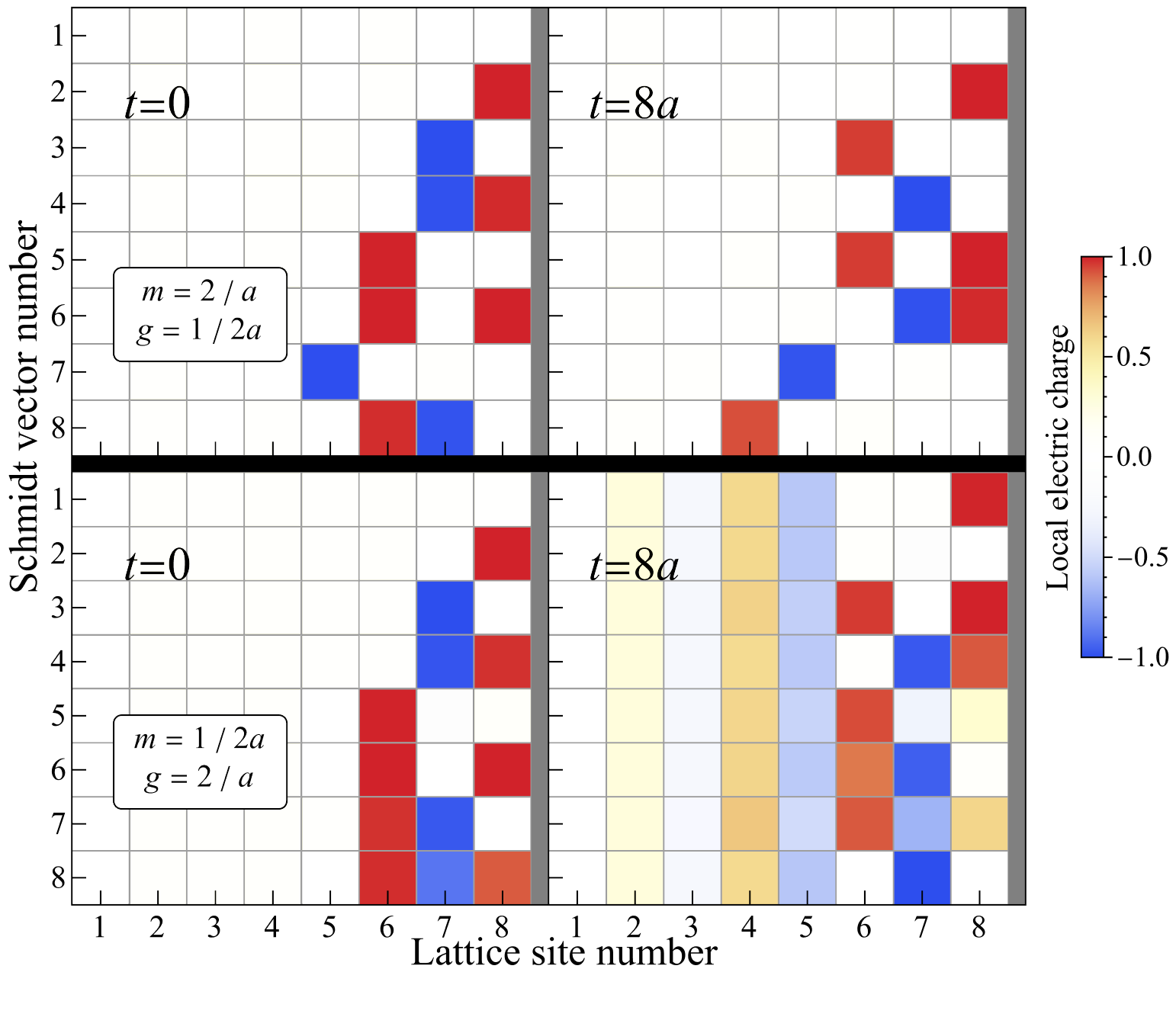}
\caption{Evolution of the electric charge distribution in the leading 8 Schmidt vectors on the lattice of size $N=16$. Top left: initial time $t=0$ with $m=2/a$ and $g=1/(2a)$; top right: time near the end of the simulation, $t=8a$ with $m=2/a$ and $g=1/(2a)$. Bottom: the same but with $m=1/(2a)$ and $g=2/a$. Horizontal direction of the grid spans half the lattice, vertical direction spans 8 leading Schmidt vectors. Schmidt vectors are ordered top to bottom according to their eigenvalues at the corresponding moment in time, with the largest eigenvalue at the top.}
\label{fig:charge_dist}
\end{figure} 

\begin{figure}
\includegraphics[width=\linewidth]{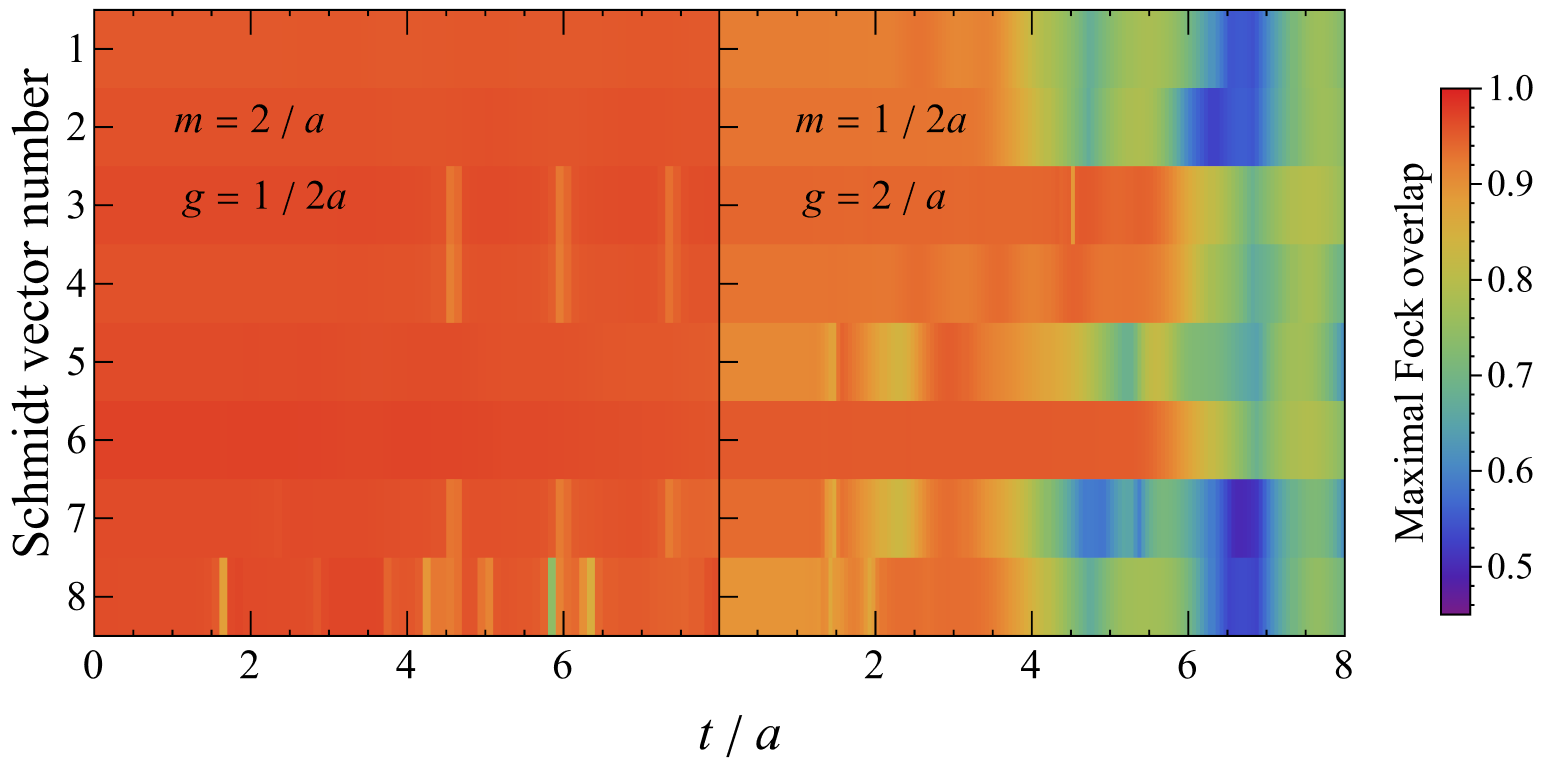}
\caption{Maximal overlap of each Schmidt vector with any Fock state. Comparison between $m=2/a, g= 1/(2a)$ on the left panel and $m=1/(2a), g=2/a$ on the right panel is shown. In both cases, $N=16$. To study continuous evolution, we choose to consider the 8 leading Schmidt vectors in the vacuum state at $t=0$ and follow their evolution. Because of the level crossing in Schmidt spectrum, at later times these vectors are not necessarily the 8 leading Schmidt vectors.}
\label{fig:Schmidt_Fock}
\end{figure}

\section{Partons and Fock states} \label{sec:Fock}
What is the physical meaning of the states in the Schmidt basis, $|\psi_i^L\rangle$?  With exact diagonalization we have direct access to the states in Schmidt basis. In order to elucidate the relationship between the Schmidt vectors and partonic states (with the fundamental fermions of the Schwinger model playing the role of partons) we compare the properties of the Schmidt vectors in two regimes of the theory: at weak coupling and at strong coupling. Throughout this section we study the system with lattice size $N=16$ and two sets of mass and the coupling constant: i) $m=2/a,~ g = 1/(2a)$ and ii) $m=1/(2a),~ g=2/a$.

First, we study the electric charge distribution in the Schmidt vectors of the vacuum state ($t=0$) and in the state perturbed by the jets ($t=8a$), displayed in Fig.~\ref{fig:charge_dist}. At weak coupling (Fig.~\ref{fig:charge_dist}, top panel) the charge is strongly concentrated in one or two lattice sites, regardless of the presence of external sources. At strong coupling without the external source (Fig.~\ref{fig:charge_dist}, bottom left panel) the charge is concentrated in the particular sites, however as a result of jet propagation (Fig.~\ref{fig:charge_dist}, bottom right panel) the electric charge gets distributed in multiple sites. In the partonic language it means that the Schmidt vectors resulting from jet fragmentation are much closer to the partonic states at weak coupling than at strong coupling. Moreover, strong coupling implies a stronger screening of the electric field produced by the external sources. This screening is manifest if one compares the top and bottom right panels of Fig.~\ref{fig:charge_dist} showing the late-time charge distribution. At strong coupling (Fig.~\ref{fig:charge_dist}, bottom panel) the leading Schmidt vectors have positive charge (thus contributing to screening) or at least charge zero. On the other hand, at weak coupling (Fig.~\ref{fig:charge_dist}, top panel) there are negative-charge states (that contribute to anti-screening) relatively high in the Schmidt spectrum.

At infinite fermion mass $m$ and in the occupation number basis, the vacuum state of our system is the so-called N\'eel state: $|\mathcal{N}\rangle = |1010\dots10\rangle$ (which translates into unoccupied fermion and antifermion sites). The N\'eel state is a useful zeroth approximation to the true vacuum state even for the finite values of the fermion mass used in this study. Deviations from the N\'eel state can be classified according to which pairs of sites (one for fermion and another one for antifermion) are ``excited" with respect to the N\'eel state. The simplest example is given by ``one-pair" excitations  $|ij\rangle = \chi_i \chi^\dagger_j |\mathcal{N}\rangle$, where $i$ is odd and $j$ is even. In the large mass and infinite volume limits $|ij\rangle$ would be the $(N^2/4)$-fold degenerate first excited states above the vacuum N\'eel state. Note that a single excitation is prohibited in the charge-zero sector. Since the ground state is in this sector and the Hamiltonian evolution conserves the total charge, the state~\eqref{eq:state} is always at the charge-zero sector. All possible excitations form a basis which we will refer to as the Fock basis.

Each of the Schmidt vectors can also be decomposed in the Fock basis on half of the lattice. In this case Fock states are not restricted to the charge-zero sector, but have the same charge as the corresponding Schmidt vector. For each Schmidt vector, let us concentrate on the Fock state with which it has the largest overlap. In Fig.~\ref{fig:Schmidt_Fock} the time evolution of this maximal overlap for the 8 leading Schmidt vectors is shown in weak coupling and strong coupling regimes. At weak coupling (Fig.~\ref{fig:Schmidt_Fock}, left panel) the overlap stays close to 1 at all times, signaling that the Schmidt vectors are very close to the partonic Fock states. On the other hand, at strong coupling (Fig.~\ref{fig:Schmidt_Fock}, right panel) this overlap is also large in the vacuum state ($t=0$) but decreases during the time evolution. This is a manifestation of hadronization in real time: at first the system is described by partonic (fermionic) degrees of freedom but with jet fragmentation the relevant degrees of freedom become bosonic.

\begin{figure}
\includegraphics[width=\linewidth]{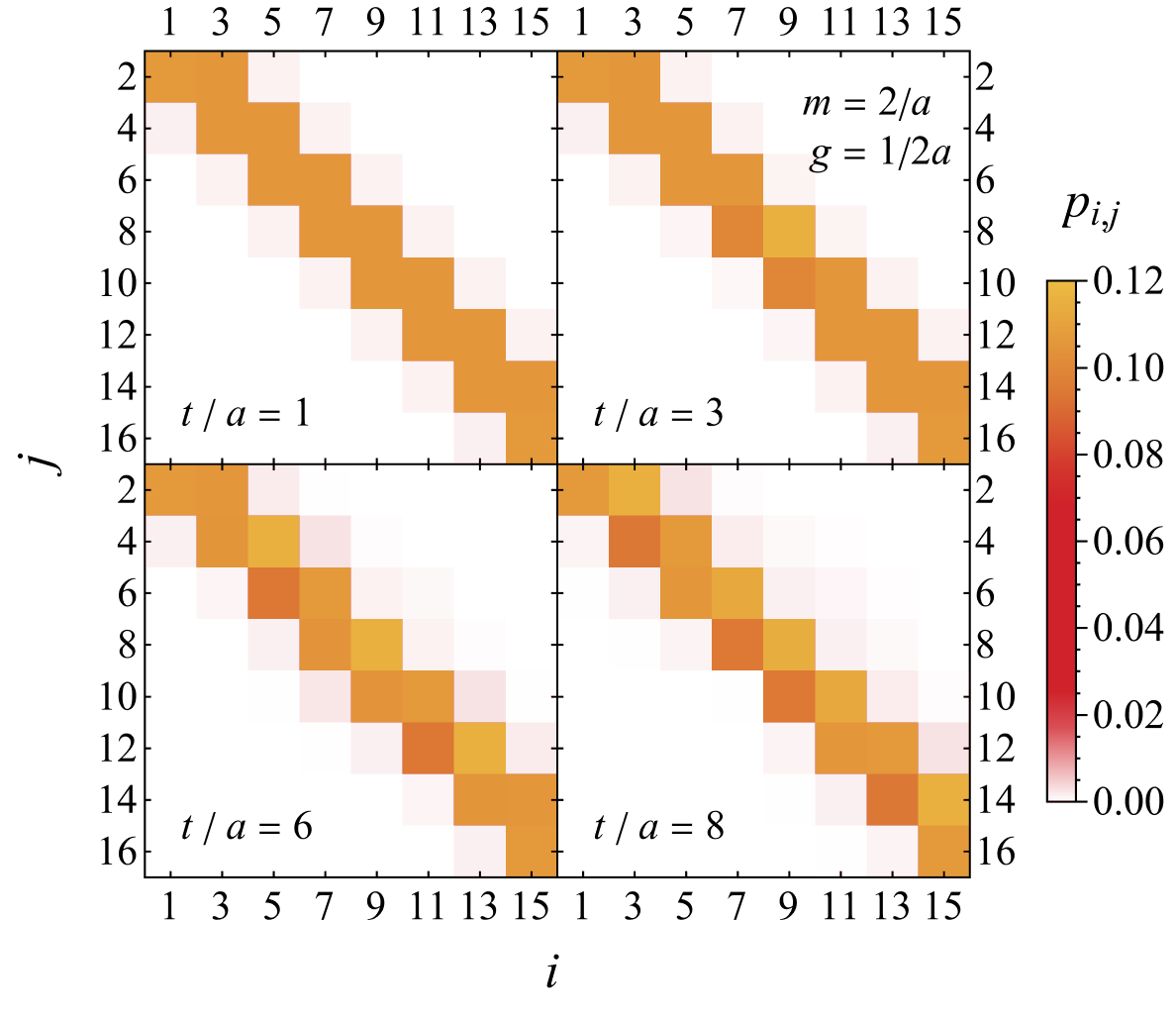}
\caption{One-pair overlaps for $N=16$, $m= 2/a, g = 1/(2a)$ for the select moments of time evolution from $t/a = 1$ to $t/a = 8$. Color denotes the value of $p_{ij}$ (Eq. \ref{eq:pair_overlap}) while the horizontal and vertical directions on the grid span $i$ and $j$, respectively.}
\label{fig:1pair_overlap_m2}
\end{figure}

\begin{figure}
\includegraphics[width=\linewidth]{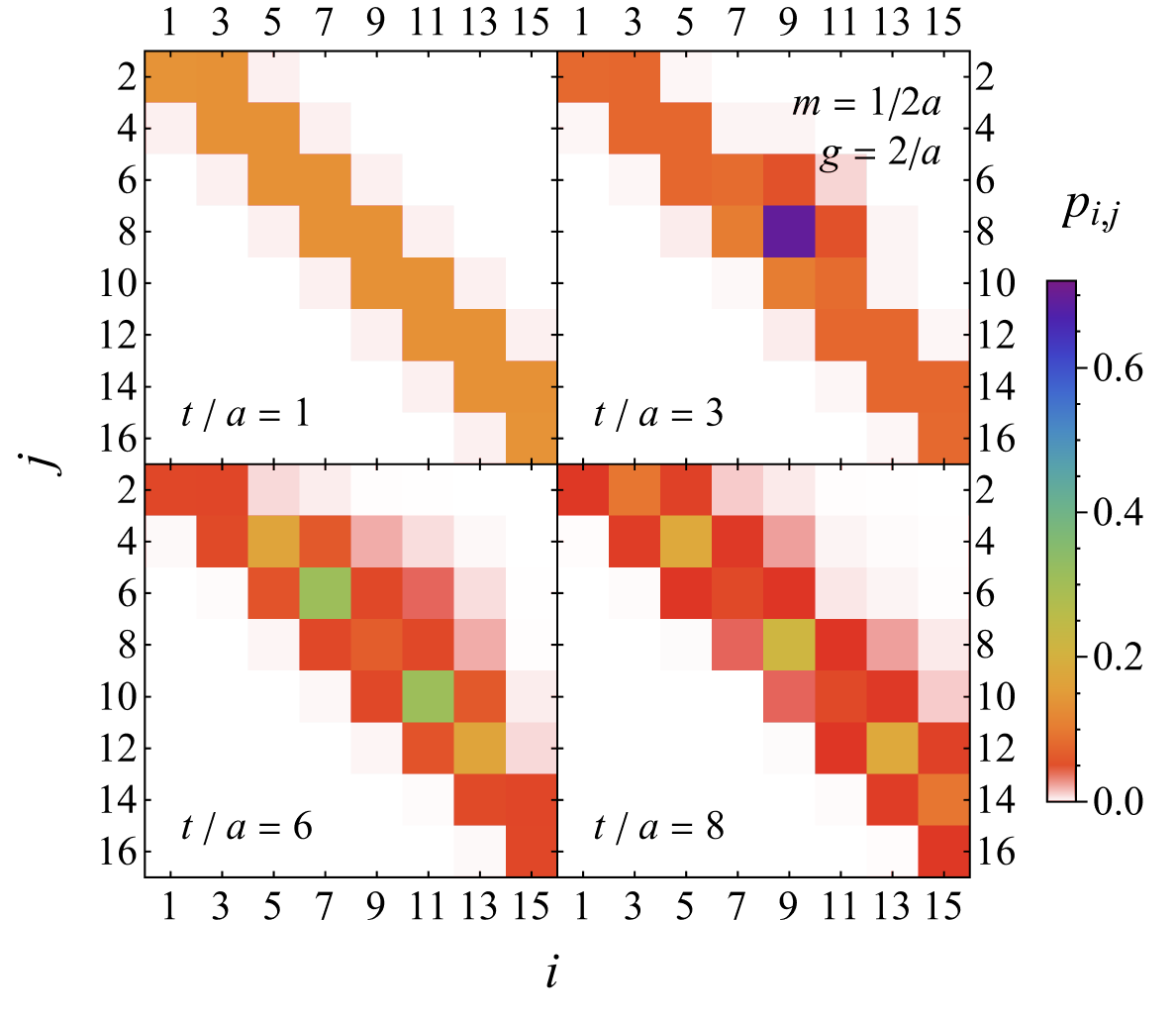}
\caption{Same as \protect{Fig.~\ref{fig:1pair_overlap_m2}} but for parameters $m=1/(2a)$ and $g=2/a$.
}
\label{fig:1pair_overlap_m05}
\end{figure}

Finally, we visualize the full quantum state evolution by calculating the probability of finding the system in one of the one-pair states:
\beq
p_{ij}^2 = |\langle\Psi_t|ij\rangle|^2.
\label{eq:pair_overlap}
\eeq
The evolution of $p_{ij}$ for two different sets of parameters is displayed in Figs.~\ref{fig:1pair_overlap_m2} and \ref{fig:1pair_overlap_m05}. In the initial state all the nearest-neighbor pairs are excited almost equally, modulo the small boundary effects. In the subsequent evolution the excitation above the vacuum state spreads along the lightcone. At weak coupling (Fig.~\ref{fig:1pair_overlap_m2}) the excitations still remain mostly nearest-neighbor but some particular pairs get excited stronger than others. At strong coupling (Fig.~\ref{fig:1pair_overlap_m05}) some next-to-nearest-neighbor pairs get substantially excited later in the evolution, while still following the lightcone propagation. This effect demonstrates that the excitations at weak coupling and larger fermion mass are more localized than in the case of strong coupling and smaller fermion mass. Moreover, at strong coupling the pairs of a certain orientation ($i>j$) are strongly preferred at later times. These are the states that screen the external electric field inside them. This is a clear indication that the screening of external electric field is stronger at strong coupling.

\section{Vacuum modification and approach to thermalization} 

The propagating jets can modify the nonperturbative structure of the vacuum. We quantify this modification by measuring the local scalar fermion density $\langle \bar\psi \psi \rangle$ that is known to have a negative value in the vacuum of the Schwinger model. On the lattice it is related to the local charge density as:
\begin{equation} \label{eq:loc_cond}
\nu_{n,t} \equiv \frac{(-1)^n}{a} q_n - \frac{1}{2a}.
\end{equation}

\begin{figure}
\includegraphics[width=1.15\linewidth]
{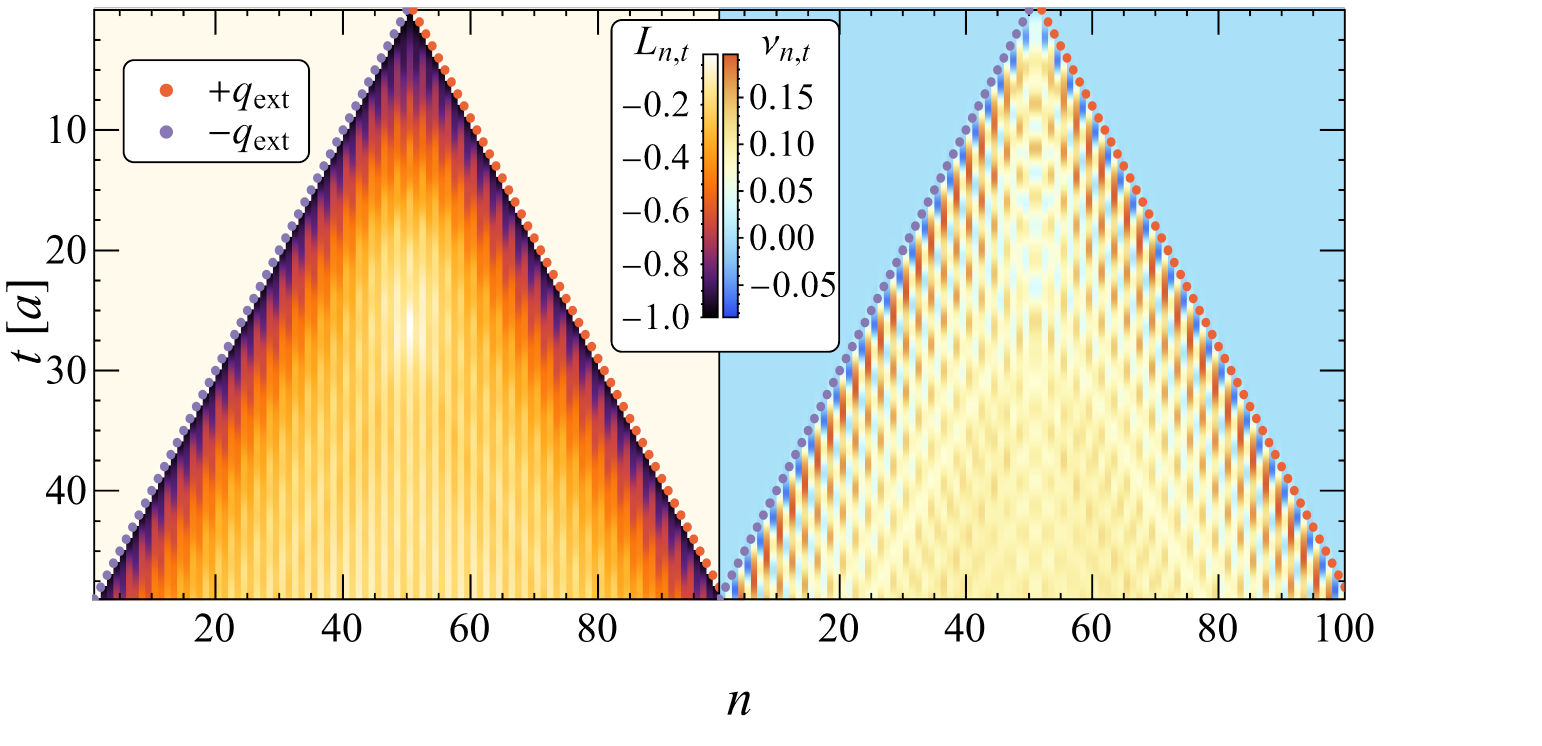}
\caption{Left panel: The distribution of electric field (\ref{eq:loc_Efield}) as a function of space and time. The vacuum value of the electric field is subtracted. Right panel: the same for the local fermion condensate (\ref{eq:loc_cond}). The system parameters are $m=1/(4a), g = 1/(2a), N = 100$. Red and blue circles denote the positions of the external positive and negative charge respectively at the corresponding time.}
\label{fig:spacetimeN100}
\end{figure}

In the presence of external sources this local condensate deviates from its vacuum value due to jet propagation. This deviation as a function of both space and time for $N=100, m = 1/(4a), g = 1/(2a)$ is shown in the right panel of Fig.~\ref{fig:spacetimeN100}. As one can see, the condensate disruption closely follows the propagating jets. At late times the value of the condensate settles to a nearly constant value (different from the vacuum one) in the central region. To see how exactly this relaxation to a new value happens, we average the deviation of the condensate from its vacuum value over a  region of size $L$ in the center of the system:
\beq \label{eq:avg_cond}
\nu_C(L)  \equiv \frac{1}{L}\sum_{n=N/2-L/2+1}^{N/2+L/2} (\nu_{n,t} - \nu_{n,0}).
\eeq

\begin{figure}
\includegraphics[width=\linewidth]{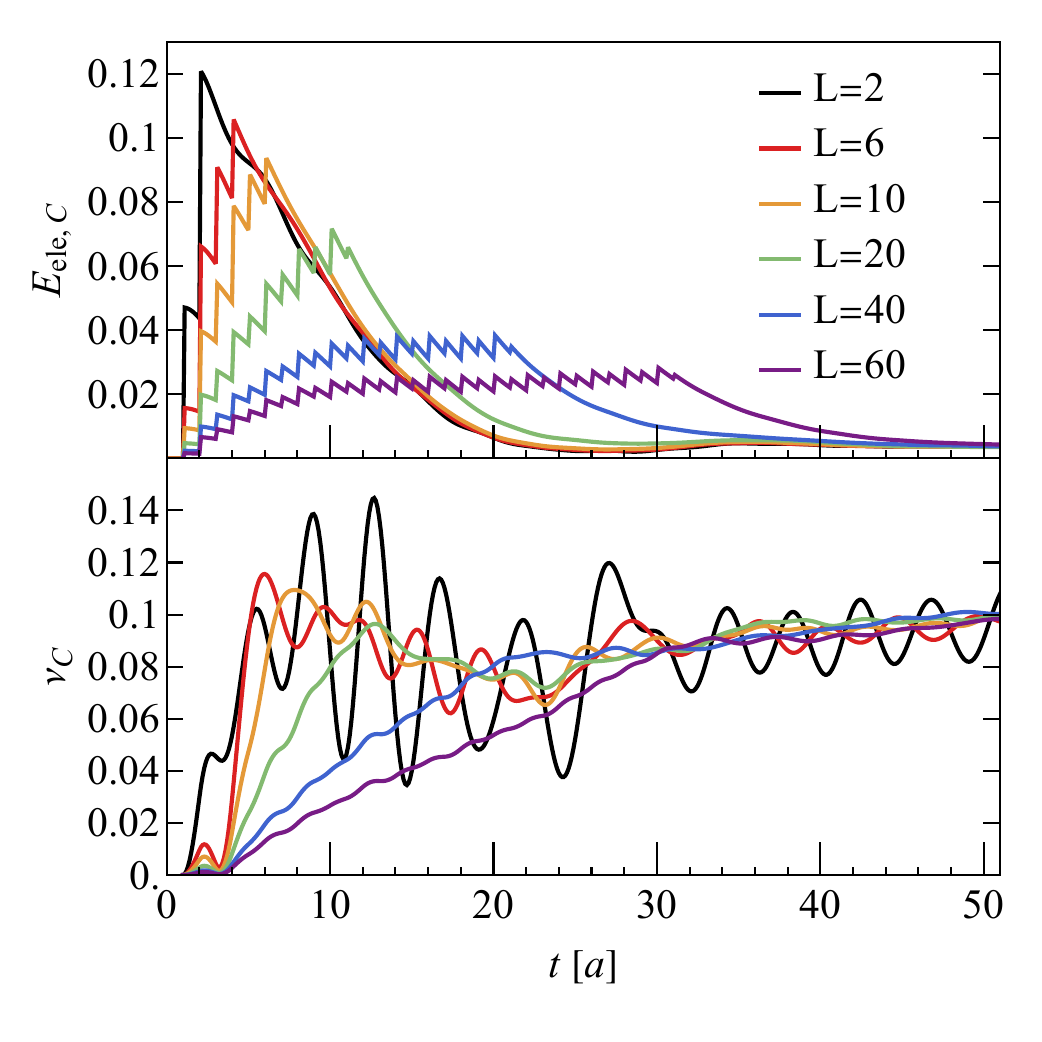}
\caption{Top panel: averaged electric field energy (\ref{eq:avg_eleE}) modulo its vacuum value as a function of time. Bottom panel: the same for fermion condensate (\ref{eq:avg_cond}). The system parameters are $m=1/(4a), g = 1/(2a), N = 100$.}
\label{fig:obs_thermalization}
\end{figure}

In Fig.~\ref{fig:obs_thermalization}, bottom panel, we show $\nu_C(L)$ as a function of time for a few different values of the central region size. We find that at sufficiently late times the average value of the condensate reaches the same constant value regardless of the size of the region where averaging is performed. This signals a step towards thermalization induced by the jets in the central part of the system.

We also study screening of the electric field produced by the external sources. The total electric field at link $n$ is given by the sum of dynamical and external fields:

\beq \label{eq:loc_Efield}
L_{n,t} = \langle \Psi_t|\Ldyn{n}|\Psi_t\rangle + \langle \Psi_t|\Lext{n}|\Psi_t\rangle,
\eeq
with $\Ldyn{n}$ and $\Lext{n}$ given by Eqs.~\eqref{eq:E_dyn} and \eqref{eq:E_ext}, respectively. In the left panel of Fig.~\ref{fig:spacetimeN100} we show the space and time dependence of the electric field, again in terms of its deviation from the vacuum value. The dynamics of external field screening is very apparent: at first, the electric field at position $n$ experiences a sudden jump when the external charge moves past $n$. Then, the electric field relaxes to a nearly constant equilibrium value achieved everywhere except the very edges of the system. We also investigate the thermalization aspect of electric field behavior in the central region. To do so, we find the average electric energy over an interval of length $L$ at the center:
\beq \label{eq:avg_eleE}
E_{\text{ele, C}}(L) = \frac{1}{L}~ \frac{a g^2}{2}\sum_{n=N/2-L/2+1}^{N/2+L/2} (\Ldyn{n}+\Lext{n}(t))^2.
\eeq
In Fig.~\ref{fig:obs_thermalization}, top panel, the deviation of the averaged electric field energy from its vacuum value is shown as a function of time for several values of the central region size. At late times we again see the signs of thermalization in the central region.

\section{Area and volume law of entanglement}

While suggestive, the previous section's results only show that the system locally equilibrates to some non-trivial steady state.  To look into the problem of thermalization further, in this section we scrutinize the behavior of entanglement as a function of time and subsystem size.

Ground states of gapped theories are expected to display an $area$ law for entanglement entropy; see Ref.~\cite{Eisert:2008ur} for a comprehensive review. On the other hand, typical quantum states, as well as thermal states, follow a volume law, see for instance~\cite{Bianchi:2021aui}. Indeed, the thermodynamical entropy is an extensive quantity. As a result, exhibiting a transition from area to volume law is crucial for establishing thermalization.

To address this question, we look at the entanglement of central subregions with the rest of the lattice, see Fig.~\ref{fig:entropy_AB}, panel (b). Computing the entanglement entropy of a MPS between two subregions that are not separated by a single site is hard, as one needs to fully reconstruct the reduced density matrix of these subregions (in the case of a single interface, the information is contained in the singular values at the interface, which are by construction easily accessible from a MPS). On the other hand, low moments of the density matrix can be reconstructed with only polynomial scaling in the bond dimension, allowing for an efficient computation. As a result, we investigate the behavior of the second-order R\'enyi entropy $S_2=-\ln[\mathrm{Tr}(\rho_A^2)]$, which is also expected to transition between an area law and a volume law~\cite{Nakagawa:2017yiw}. 

\begin{figure}
    \centering
    \includegraphics[scale=0.5]{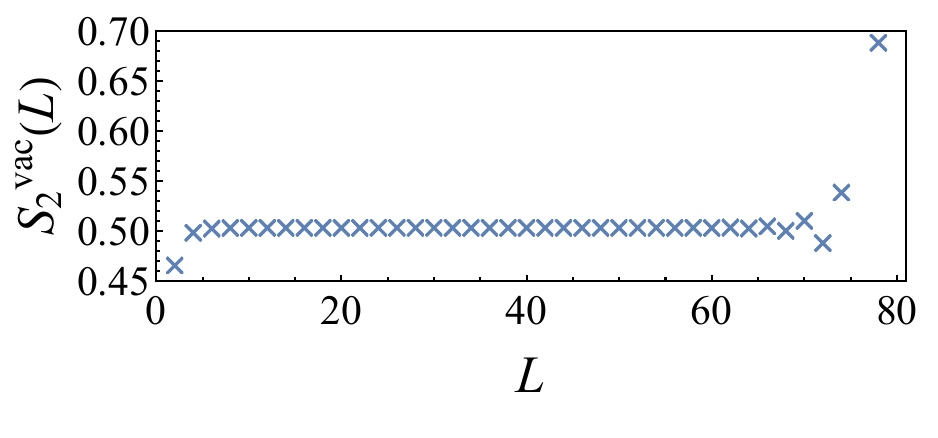}
    \caption{Second R\'enyi entropy in the vacuum, at $t=0$.}
    \label{fig:S2_vac}
\end{figure}

As a reference point, we present in Fig.~\ref{fig:S2_vac} its value as a function of the subsystem size $L$, in the vacuum state ($t=0$). The system size for this investigation is $N=80$, as the growth in the bond dimension as a function of time impeded us from computing the R\'enyi entropy for time larger than $40$ (and thus did not warrant the use of a lattice larger than $N=80$). We see that, except for very small ($L=2,4$) and very large ($L>70$, close to the boundaries) subsystem sizes, the second R\'enyi entropy is independent of the subsystem size and as a consequence does follow an area law. 

\begin{figure}
    \centering
    \includegraphics[scale=0.5]{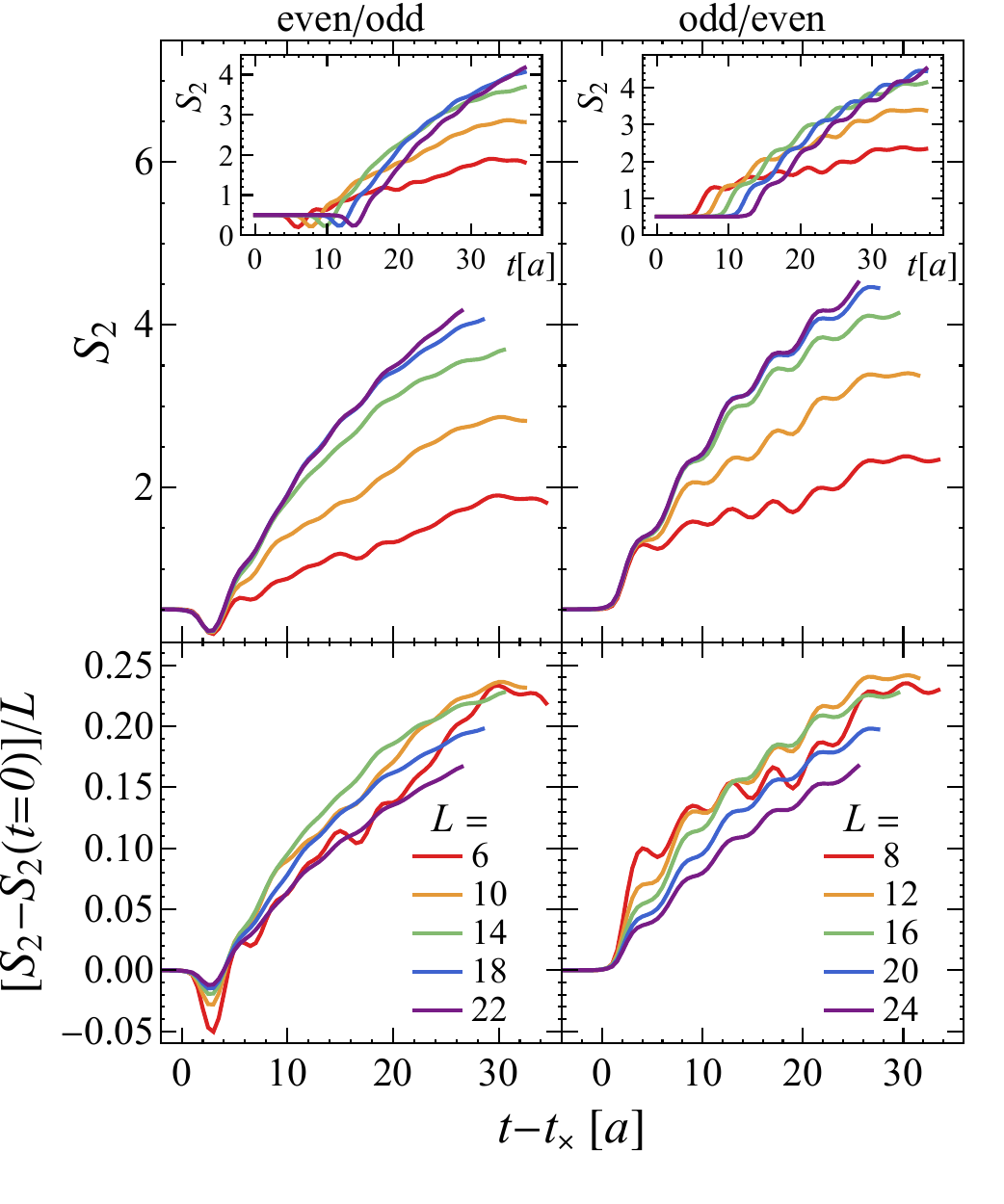}
    \caption{\textbf{Top panel}: second R\'enyi entropy as a function of $t-t_{\times}$, with $t_{\times}$ the time at which the external sources cross the boundaries of the sublattice, for different sublattice in the center. The left(right) panels correspond to odd sites-even sites(even sites-odd sites) boundaries. They differ because of the staggering, which is a lattice artifact. \textbf{Bottom panel}: substracted second Renyi entropy rescaled by the size of the sublattice as a function of $t-t_{\times}$ for different sublattice in the center. Smaller subsystems start exhibiting volume law at a late time, exhibited by the fact the saturate to the same value on this plot. The left and right panels correspond to odd sites- even sites and even sites-odd sites boundaries, respectively. \textbf{Insets}: the corresponding second Renyi entropy as a function of time.}
    \label{fig:S2_master}
\end{figure}

The time-dependence of R\'enyi entropies for several different values of $L$ is shown in Fig.~\ref{fig:S2_master}. For readability, we separated the cases with $L/2$ odd (corresponding to even and odd sites at the boundary) and even (corresponding to odd and even sites at the boundary). While this distinction is irrelevant in the continuum limit, the staggering induces quantitative differences. The qualitative features, on the other hand, are the same. As soon as the external particles cross the boundaries of the subsystem ($t=t_\times = L/2$) the R\'enyi entropies start growing. Initially, the growth is independent of the subsystem size; the state still follows an area law. After some time which grows with the subsystem size, the value of the entanglement entropy starts depending on the subsystem size, before eventually reaching a plateau (in the present simulation the plateau is reached only for $L\in[6,16]$). Moreover, the leading dependence on the system size is linear once the plateau is reached. This is demonstrated in the lower panel of Fig.\ref{fig:S2_master}, where we show the value of the R\'enyi entropy with the vacuum value subtracted, divided by the subsystem size. In other words, the system transitions from following the area law to the volume law of entanglement.

We note that, even if in a different setup, the observed behavior is reminiscent of the typical response of many-body ground states to global quenches. There, subsystems in finite volumes follow an area law as a function for a finite time $t_a$, before transitioning to a volume law once relaxed. Moreover, at early time $t<t_a$, the value of the entanglement entropy is often seen to grow linearly as a function of time (irrespectively of the subsystem size, as a consequence of the area law) \cite{Eisert:2008ur}. This seems also to be the case in our setup. 

We conclude this section with another observation on the dynamics of the entanglement entropy. As we just observed, the entanglement entropy follows an area law for subsystems that have not had time to equilibrate yet. 
This can be rephrased in the following way:  the degrees of freedom responsible for entanglement between the two subsystems reside within few units of the correlation length from the boundary between the subsystems. 
This is supported by the fact already discussed in Section \ref{sec:Fock}, where we showed that the electric charge distribution in the leading Schmidt vectors is mostly localized near the boundary.

In our dynamical setup this realization of the area law has interesting consequences. In particular, the external charges can only affect the entanglement between subregions after a finite time set by the distance from the charges to the nearest entanglement surface (i.e. point in one dimension). This is a relativistic equivalent to the Lieb-Robinson speed of information~\cite{Eisert:2008ur}, which is bounded by the speed of light in our model. 

\begin{figure}
\includegraphics[scale=0.45]{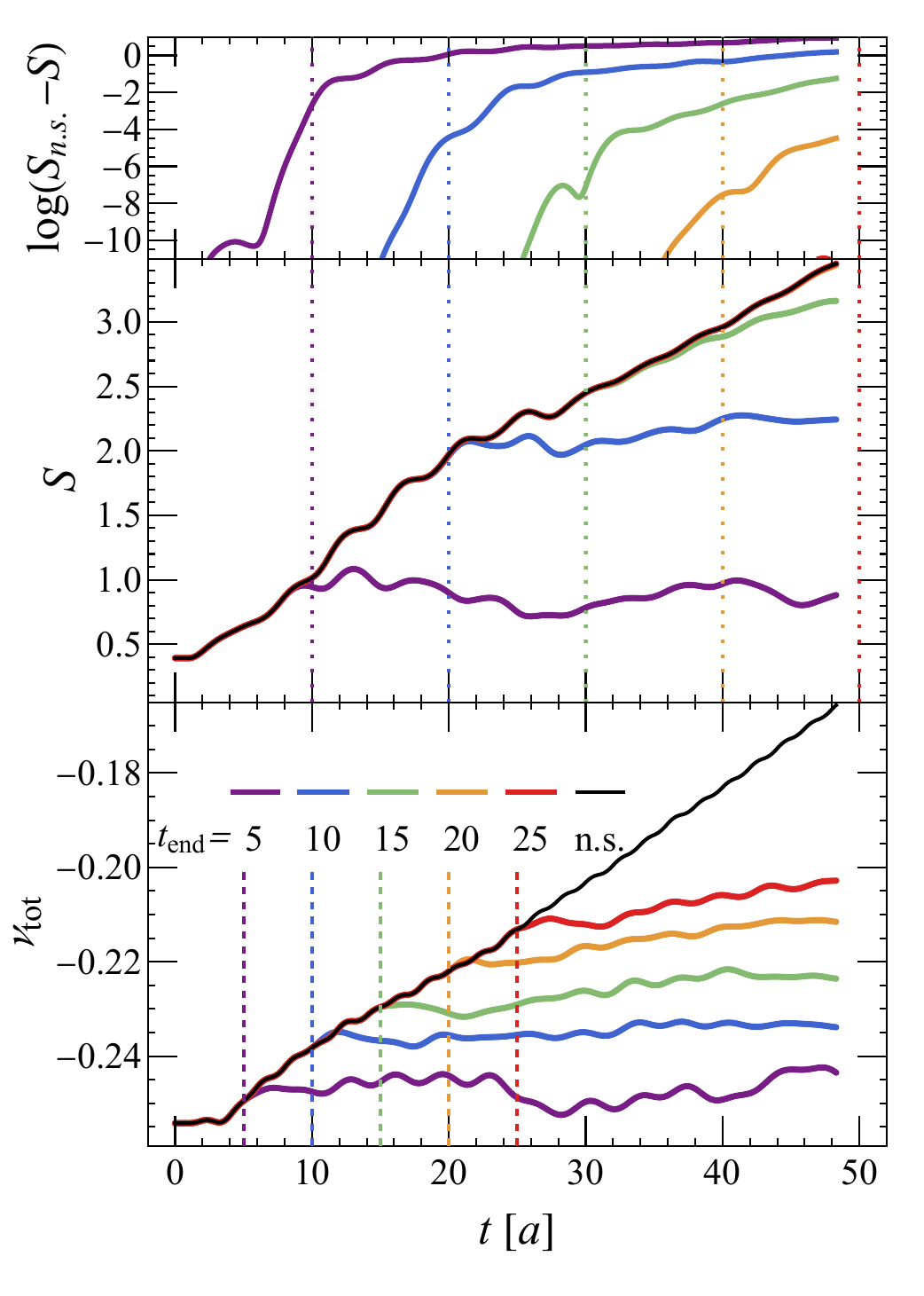}
\caption{Middle panel: the evolution of entanglement entropy with the external source describing stopped jets, Eq.(\ref{eq:jet_stop}) for different values of $t_{end}$ or non-stopped jets (n.s.). Top panel: the logarithm of the difference between the entanglement entropy in the non-stopped case ($S_{\rm{n.s.}}$) and the entanglement entropy in each of the stopping scenarios. Bottom panel: the evolution of the integrated chiral condensate with the external source describing stopped jets for different values of $t_{end}$. The parameters of the system are $N=100, m = 1/(4a), g = 1/(2a)$. Vertical dashed lines highlight $t_{end}$ when the jets are stopped for each corresponding curve. Vertical dotted lines highlight $2t_{end}$.}
\label{fig:jet_stop}
\end{figure}

To verify this expectation, we consider again the bipartite entanglement at the middle of the lattice. We modify our setup so that the jets are stopped after a certain time $t_{end}$. Concretely, the external electric field becomes as follows instead of eq.~\eqref{eq:E_ext}:

\beq
\Lext{n}(t) = \begin{cases} 
-\theta\left(\frac{t-t_0}{a} - \left|n-\frac{N}{2}\right|\right), &  t\leq t_{end} \\ 
-\theta\left(\frac{t_{end}-t_0}{a} - \left|n-\frac{N}{2}\right|\right), & t > t_{end}  \\
\end{cases} \label{eq:jet_stop}
\eeq

For the system size $N=100$ we compare the evolution of entanglement entropy for $t_{end}=5$, $t_{end}=10$, $t_{end}=15$ $t_{end}=20$ and $t_{end}=25$ (for brevity, from now on the time is measured in the units of $a$) and the original setup without stopping the jets until they reach the end of the lattice. The results are displayed in Fig. \ref{fig:jet_stop}, middle panel. As one can see, if $t_{end} = 5$, entanglement entropy evolution until $t\approx 10$ is consistent with the one where jets are not stopped. This means that the signal about the jets being stopped propagates back to the center of the lattice at a speed close to the speed of light. The evolution of entanglement entropy for $t_{end} = 10$ and $t_{end} = 15$ follows the same trend. For $t_{end} = 20$ the deviation from the non-stopped case becomes too small but it is still present, as can be seen in the logarithmic scale in the top panel of Fig. \ref{fig:jet_stop}. As one can see, the deviation from the non-stopped case becomes significantly smaller at larger $t_{end}$. It is another manifestation of the fact that the most important contribution to the entanglement entropy comes from the degrees of freedom localized near the interface between the subsystems.

To be contrasted with the evolution of the entanglement entropy, in Fig.~\ref{fig:jet_stop}, bottom panel, we show the evolution of a local observable, the chiral condensate integrated over the whole system, in the same setup. We find that a local observable responds to a change in the quench dynamics almost instantaneously, with only a small delay independent of the system's size. This behavior is expected from an extensive local observable that obeys the volume law rather than the area law. 

\section{Summary and outlook}
In this work, we investigated the growth of entanglement and the approach to thermalization of the propagating jets in the $1+1$-dimensional massive Schwinger model. We vastly expanded on the results of ~\cite{Florio:2023dke} by evaluating the entanglement spectrum, and by using tensor network techniques to access larger lattices and longer jet evolution. We refined the understanding of entanglement generation in the system by studying the symmetry-resolved half-lattice Schmidt spectrum. Combining these results with exact diagonalization, we achieved a qualitative understanding of the asymptotic states of the theory. 

We also studied as a function of the coupling the nature of the asymptotic states that contribute to the entanglement entropy. At weak coupling, these states are very close to partonic Fock states. On the other hand, at  strong coupling, the overlap between the Schmidt states and the partonic states drops at late time, likely signaling a transition towards bosonic Fock states. In other words, we directly witnessed, in real-time, the equivalent of hadronization in the massive Schwinger model. 

Next, we observed that the central rapidity region of our system relaxes to a steady state at late times and that the steady state is independent of the subsystem's size. This suggests that particle production leads to  thermalization. While this can only be strictly assessed by comparing all properties of this steady state to thermal ensembles, something we leave for future works, the thermalization conjecture is further reinforced by the behavior of entanglement. By computing the second R\'enyi entropy over central regions of the system, we could clearly observe a transition between the area law for entanglement and the volume law, which is necessary to obtain an emergent extensive thermodynamical entropy. We concluded our analysis of the entanglement entropy properties by measuring the information propagation speed. We confirmed that this speed is saturated and equals the speed of light.

At a time when direct experimental measurements of entanglement, not only in deep-inelastic scattering \cite{Hentschinski:2023izh}, but also in high-energy proton-proton collisions  (see e.g. the recent measurement by the ATLAS collaboration \cite{ATLAS:2023fsd}) become possible, stepping up first principle, real-time investigations of entanglement structure and entanglement dynamics in relativistic field theory is of utmost importance. This work is a step in this direction; it opens fascinating directions for future research. The first one is to characterize thermalization more precisely. In particular, as the system is closed and the evolution is one of a pure state, apparent thermalization is directly reminiscent of the Eigenstate Thermalization Hypothesis (ETH), which is the lore in many-body physics regarding quantum thermalization (for review see e.g. \cite{Deutsch:2018ulr}).  However, ETH deals with finite-size systems. Making this relation more concrete or even extending the ETH to this dynamical setting, is of great interest and would help to bridge the growing body of literature on thermalization in gauge theory from the quantum dynamics, see e.g. \cite{Kharzeev:2005iz, Berges:2017hne,Zhou:2021kdl, Mueller:2021gxd, Desaules:2022ibp,Ebner:2023ixq,Yao:2023pht, Grieninger:2023ufa}. The connection to many-body physics and quantum information theory also offers a variety of other exciting quantities to measure. For instance, the behavior of information scrambling in this model is interesting to investigate; it is  expected to be related to the saturation time of our second R\'enyi entropy \cite{Nakagawa:2017yiw}. The behavior of mixed state entanglement would also be interesting to investigate  \cite{Florio:2023mzk}.
\vskip0.3cm 

To summarize, we have investigated the real-time behavior of entanglement entropy and entanglement spectrum in massive Schwinger model coupled to external sources -- a setup that mimics the production of jets in high energy QCD. Our results provide an intriguing insight into the microscopic dynamics of hadronization and thermalization in confining quantum field theories, and suggest new directions for theoretical and experimental research. 

\vskip0.3cm

\section*{Acknowledgement}
We thank Swagato Mukherjee for useful discussions and communications. We acknowledge valuable discussions at the workshop ``Thermalization, from Cold Atoms to Hot Quantum Chromodynamics" that was held during September 2023 at the InQubator for Quantum Simulation (IQuS) hosted by the Institute for Nuclear Theory (INT). The InQubator for Quantum Simulation is supported by U.S. Department of Energy, Office of Science, Office of Nuclear Physics, InQubator for Quantum Simulation (IQuS) under Award Number DOE (NP) Award DE-SC0020970 via the program on Quantum Horizons: QIS Research and Innovation for Nuclear Science. AF is particularly grateful to Andreas Elben and Torsten Zache for insightful explanations on R\'enyi entropies and Mark Srednicki and Anatoli Polkovnikov on the ETH. DK is grateful to Yuri Dokshitzer for many discussions on the possible origin of the local parton-hadron duality. 

This work was supported by the U.S. Department of Energy, Office of Science, National Quantum Information Science Research Centers, Co-design Center for Quantum Advantage (C2QA) under Contract No.DE-SC0012704 (AF, KI, DK, VK), the U.S. Department of Energy, Office of Science, Office of Nuclear Physics, Grants Nos. DE-FG88ER41450 (DF, DK, SS) and DE-SC0012704 (AF, DK, KY), and Tsinghua University under grant No. 53330500923 (SS). 
This research used resources of the National Energy Research Scientific Computing Center, a DOE Office of Science User Facility supported by the Office of Science of the U.S. Department of Energy under Contract No. DE-AC02-05CH11231 using NERSC award NERSC DDR-ERCAP0028999. 
\clearpage
\appendix 
\bibliographystyle{utphys}
\bibliography{main}

\providecommand{\href}[2]{#2}\begingroup\raggedright\begin{thebibliography}{10}

\bibitem{Kharzeev:2017qzs}
D.~E. Kharzeev and E.~M. Levin, ``{Deep inelastic scattering as a probe of
  entanglement},'' \href{http://dx.doi.org/10.1103/PhysRevD.95.114008}{{\em
  Phys. Rev. D} {\bfseries 95} no.~11, (2017) 114008},
  \href{http://arxiv.org/abs/1702.03489}{{\ttfamily arXiv:1702.03489
  [hep-ph]}}.

\bibitem{Tu:2019ouv}
Z.~Tu, D.~E. Kharzeev, and T.~Ullrich, ``{Einstein-Podolsky-Rosen Paradox and
  Quantum Entanglement at Subnucleonic Scales},''
  \href{http://dx.doi.org/10.1103/PhysRevLett.124.062001}{{\em Phys. Rev.
  Lett.} {\bfseries 124} no.~6, (2020) 062001},
  \href{http://arxiv.org/abs/1904.11974}{{\ttfamily arXiv:1904.11974
  [hep-ph]}}.

\bibitem{Hentschinski:2023izh}
M.~Hentschinski, D.~E. Kharzeev, K.~Kutak, and Z.~Tu, ``{Probing the Onset of
  Maximal Entanglement inside the Proton in Diffractive Deep Inelastic
  Scattering},'' \href{http://dx.doi.org/10.1103/PhysRevLett.131.241901}{{\em
  Phys. Rev. Lett.} {\bfseries 131} no.~24, (2023) 241901},
  \href{http://arxiv.org/abs/2305.03069}{{\ttfamily arXiv:2305.03069
  [hep-ph]}}.

\bibitem{H1:2020zpd}
{\bfseries H1} Collaboration, V.~Andreev {\em et~al.}, ``{Measurement of
  charged particle multiplicity distributions in DIS at HERA and its
  implication to entanglement entropy of partons},''
  \href{http://dx.doi.org/10.1140/epjc/s10052-021-08896-1}{{\em Eur. Phys. J.
  C} {\bfseries 81} no.~3, (2021) 212},
  \href{http://arxiv.org/abs/2011.01812}{{\ttfamily arXiv:2011.01812
  [hep-ex]}}.

\bibitem{Kharzeev:2021nzh}
D.~E. Kharzeev, ``{Quantum information approach to high energy interactions},''
  \href{http://dx.doi.org/10.1098/rsta.2021.0063}{{\em Phil. Trans. A. Math.
  Phys. Eng. Sci.} {\bfseries 380} no.~2216, (2021) 20210063},
  \href{http://arxiv.org/abs/2108.08792}{{\ttfamily arXiv:2108.08792
  [hep-ph]}}.

\bibitem{Armesto:2019mna}
N.~Armesto, F.~Dominguez, A.~Kovner, M.~Lublinsky, and V.~Skokov, ``{The Color
  Glass Condensate density matrix: Lindblad evolution, entanglement entropy and
  Wigner functional},'' \href{http://dx.doi.org/10.1007/JHEP05(2019)025}{{\em
  JHEP} {\bfseries 05} (2019) 025},
  \href{http://arxiv.org/abs/1901.08080}{{\ttfamily arXiv:1901.08080
  [hep-ph]}}.

\bibitem{Kharzeev:2021yyf}
D.~E. Kharzeev and E.~Levin, ``{Deep inelastic scattering as a probe of
  entanglement: Confronting experimental data},''
  \href{http://dx.doi.org/10.1103/PhysRevD.104.L031503}{{\em Phys. Rev. D}
  {\bfseries 104} no.~3, (2021) L031503},
  \href{http://arxiv.org/abs/2102.09773}{{\ttfamily arXiv:2102.09773
  [hep-ph]}}.

\bibitem{Dvali:2021ooc}
G.~Dvali and R.~Venugopalan, ``{Classicalization and unitarization of wee
  partons in QCD and gravity: The CGC-black hole correspondence},''
  \href{http://dx.doi.org/10.1103/PhysRevD.105.056026}{{\em Phys. Rev. D}
  {\bfseries 105} no.~5, (2022) 056026},
  \href{http://arxiv.org/abs/2106.11989}{{\ttfamily arXiv:2106.11989
  [hep-th]}}.

\bibitem{Zhang:2021hra}
K.~Zhang, K.~Hao, D.~Kharzeev, and V.~Korepin, ``{Entanglement entropy
  production in deep inelastic scattering},''
  \href{http://dx.doi.org/10.1103/PhysRevD.105.014002}{{\em Phys. Rev. D}
  {\bfseries 105} no.~1, (2022) 014002},
  \href{http://arxiv.org/abs/2110.04881}{{\ttfamily arXiv:2110.04881
  [quant-ph]}}.

\bibitem{Liu:2022ohy}
Y.~Liu, M.~A. Nowak, and I.~Zahed, ``{Entanglement entropy and flow in
  two-dimensional QCD: Parton and string duality},''
  \href{http://dx.doi.org/10.1103/PhysRevD.105.114027}{{\em Phys. Rev. D}
  {\bfseries 105} no.~11, (2022) 114027},
  \href{http://arxiv.org/abs/2202.02612}{{\ttfamily arXiv:2202.02612
  [hep-ph]}}.

\bibitem{Liu:2022hto}
Y.~Liu, M.~A. Nowak, and I.~Zahed, ``{Rapidity evolution of the entanglement
  entropy in quarkonium: Parton and string duality},''
  \href{http://dx.doi.org/10.1103/PhysRevD.105.114028}{{\em Phys. Rev. D}
  {\bfseries 105} no.~11, (2022) 114028},
  \href{http://arxiv.org/abs/2203.00739}{{\ttfamily arXiv:2203.00739
  [hep-ph]}}.

\bibitem{Dumitru:2023qee}
A.~Dumitru, A.~Kovner, and V.~V. Skokov, ``{Entanglement entropy of the proton
  in coordinate space},''
  \href{http://dx.doi.org/10.1103/PhysRevD.108.014014}{{\em Phys. Rev. D}
  {\bfseries 108} no.~1, (2023) 014014},
  \href{http://arxiv.org/abs/2304.08564}{{\ttfamily arXiv:2304.08564
  [hep-ph]}}.

\bibitem{Azimov:1984np}
Y.~I. Azimov, Y.~L. Dokshitzer, V.~A. Khoze, and S.~I. Troyan, ``{Similarity of
  Parton and Hadron Spectra in QCD Jets},''
  \href{http://dx.doi.org/10.1007/BF01642482}{{\em Z. Phys. C} {\bfseries 27}
  (1985) 65--72}.

\bibitem{Dokshitzer:1991eq}
Y.~L. Dokshitzer, V.~A. Khoze, and S.~I. Troian, ``{On the concept of local
  parton hadron duality},''
  \href{http://dx.doi.org/10.1088/0954-3899/17/10/017}{{\em J. Phys. G}
  {\bfseries 17} (1991) 1585--1587}.

\bibitem{Hentschinski:2021aux}
M.~Hentschinski and K.~Kutak, ``{Evidence for the maximally entangled low x
  proton in Deep Inelastic Scattering from H1 data},''
  \href{http://dx.doi.org/10.1140/epjc/s10052-022-10056-y}{{\em Eur. Phys. J.
  C} {\bfseries 82} no.~2, (2022) 111},
  \href{http://arxiv.org/abs/2110.06156}{{\ttfamily arXiv:2110.06156
  [hep-ph]}}. [Erratum: Eur.Phys.J.C 83, 1147 (2023)].

\bibitem{Becattini:1997rv}
F.~Becattini and U.~W. Heinz, ``{Thermal hadron production in p p and p anti-p
  collisions},'' \href{http://dx.doi.org/10.1007/s002880050551}{{\em Z. Phys.
  C} {\bfseries 76} (1997) 269--286},
  \href{http://arxiv.org/abs/hep-ph/9702274}{{\ttfamily arXiv:hep-ph/9702274}}.
  [Erratum: Z.Phys.C 76, 578 (1997)].

\bibitem{Andronic:2008gu}
A.~Andronic, P.~Braun-Munzinger, and J.~Stachel, ``{Thermal hadron production
  in relativistic nuclear collisions: The Hadron mass spectrum, the horn, and
  the QCD phase transition},''
  \href{http://dx.doi.org/10.1016/j.physletb.2009.06.021}{{\em Phys. Lett. B}
  {\bfseries 673} (2009) 142--145},
  \href{http://arxiv.org/abs/0812.1186}{{\ttfamily arXiv:0812.1186 [nucl-th]}}.
  [Erratum: Phys.Lett.B 678, 516 (2009)].

\bibitem{Casher:1974vf}
A.~Casher, J.~B. Kogut, and L.~Susskind, ``{Vacuum polarization and the absence
  of free quarks},'' \href{http://dx.doi.org/10.1103/PhysRevD.10.732}{{\em
  Phys. Rev. D} {\bfseries 10} (1974) 732--745}.

\bibitem{Loshaj:2011jx}
F.~Loshaj and D.~E. Kharzeev, ``{LPM effect as the origin of the jet
  fragmentation scaling in heavy ion collisions},''
  \href{http://dx.doi.org/10.1142/S0218301312500887}{{\em Int. J. Mod. Phys. E}
  {\bfseries 21} (2012) 1250088},
  \href{http://arxiv.org/abs/1111.0493}{{\ttfamily arXiv:1111.0493 [hep-ph]}}.

\bibitem{Kharzeev:2012re}
D.~E. Kharzeev and F.~Loshaj, ``{Jet energy loss and fragmentation in heavy ion
  collisions},'' \href{http://dx.doi.org/10.1103/PhysRevD.87.077501}{{\em Phys.
  Rev. D} {\bfseries 87} no.~7, (2013) 077501},
  \href{http://arxiv.org/abs/1212.5857}{{\ttfamily arXiv:1212.5857 [hep-ph]}}.

\bibitem{Kharzeev:2013wra}
D.~E. Kharzeev and F.~Loshaj, ``{Anomalous soft photon production from the
  induced currents in Dirac sea},''
  \href{http://dx.doi.org/10.1103/PhysRevD.89.074053}{{\em Phys. Rev. D}
  {\bfseries 89} no.~7, (2014) 074053},
  \href{http://arxiv.org/abs/1308.2716}{{\ttfamily arXiv:1308.2716 [hep-ph]}}.

\bibitem{Florio:2023dke}
A.~Florio, D.~Frenklakh, K.~Ikeda, D.~Kharzeev, V.~Korepin, S.~Shi, and K.~Yu,
  ``{Real-Time Nonperturbative Dynamics of Jet Production in Schwinger Model:
  Quantum Entanglement and Vacuum Modification},''
  \href{http://dx.doi.org/10.1103/PhysRevLett.131.021902}{{\em Phys. Rev.
  Lett.} {\bfseries 131} no.~2, (2023) 021902},
  \href{http://arxiv.org/abs/2301.11991}{{\ttfamily arXiv:2301.11991
  [hep-ph]}}.

\bibitem{Bauer:2022hpo}
C.~W. Bauer {\em et~al.}, ``{Quantum Simulation for High-Energy Physics},''
  \href{http://dx.doi.org/10.1103/PRXQuantum.4.027001}{{\em PRX Quantum}
  {\bfseries 4} no.~2, (2023) 027001},
  \href{http://arxiv.org/abs/2204.03381}{{\ttfamily arXiv:2204.03381
  [quant-ph]}}.

\bibitem{Bauer:2023qgm}
C.~W. Bauer, Z.~Davoudi, N.~Klco, and M.~J. Savage, ``{Quantum simulation of
  fundamental particles and forces},''
  \href{http://dx.doi.org/10.1038/s42254-023-00599-8}{{\em Nature Rev. Phys.}
  {\bfseries 5} no.~7, (2023) 420--432}.

\bibitem{Klco:2018kyo}
N.~Klco, E.~F. Dumitrescu, A.~J. McCaskey, T.~D. Morris, R.~C. Pooser, M.~Sanz,
  E.~Solano, P.~Lougovski, and M.~J. Savage, ``{Quantum-classical computation
  of Schwinger model dynamics using quantum computers},''
  \href{http://dx.doi.org/10.1103/PhysRevA.98.032331}{{\em Phys. Rev. A}
  {\bfseries 98} no.~3, (2018) 032331},
  \href{http://arxiv.org/abs/1803.03326}{{\ttfamily arXiv:1803.03326
  [quant-ph]}}.

\bibitem{Farrell:2023fgd}
R.~C. Farrell, M.~Illa, A.~N. Ciavarella, and M.~J. Savage, ``{Scalable
  Circuits for Preparing Ground States on Digital Quantum Computers: The
  Schwinger Model Vacuum on 100 Qubits},''
  \href{http://arxiv.org/abs/2308.04481}{{\ttfamily arXiv:2308.04481
  [quant-ph]}}.

\bibitem{Farrell:2024fit}
R.~C. Farrell, M.~Illa, A.~N. Ciavarella, and M.~J. Savage, ``{Quantum
  Simulations of Hadron Dynamics in the Schwinger Model using 112 Qubits},''
  \href{http://arxiv.org/abs/2401.08044}{{\ttfamily arXiv:2401.08044
  [quant-ph]}}.

\bibitem{Zache:2018cqq}
T.~V. Zache, N.~Mueller, J.~T. Schneider, F.~Jendrzejewski, J.~Berges, and
  P.~Hauke, ``{Dynamical Topological Transitions in the Massive Schwinger Model
  with a $\theta$ Term},''
  \href{http://dx.doi.org/10.1103/PhysRevLett.122.050403}{{\em Phys. Rev.
  Lett.} {\bfseries 122} no.~5, (2019) 050403},
  \href{http://arxiv.org/abs/1808.07885}{{\ttfamily arXiv:1808.07885
  [quant-ph]}}.

\bibitem{Rigobello:2021fxw}
M.~Rigobello, S.~Notarnicola, G.~Magnifico, and S.~Montangero, ``{Entanglement
  generation in (1+1)D QED scattering processes},''
  \href{http://dx.doi.org/10.1103/PhysRevD.104.114501}{{\em Phys. Rev. D}
  {\bfseries 104} no.~11, (2021) 114501},
  \href{http://arxiv.org/abs/2105.03445}{{\ttfamily arXiv:2105.03445
  [hep-lat]}}.

\bibitem{deJong:2021wsd}
W.~A. de~Jong, K.~Lee, J.~Mulligan, M.~P\l{}osko\'n, F.~Ringer, and X.~Yao,
  ``{Quantum simulation of nonequilibrium dynamics and thermalization in the
  Schwinger model},'' \href{http://dx.doi.org/10.1103/PhysRevD.106.054508}{{\em
  Phys. Rev. D} {\bfseries 106} no.~5, (2022) 054508},
  \href{http://arxiv.org/abs/2106.08394}{{\ttfamily arXiv:2106.08394
  [quant-ph]}}.

\bibitem{Belyansky:2023rgh}
R.~Belyansky, S.~Whitsitt, N.~Mueller, A.~Fahimniya, E.~R. Bennewitz,
  Z.~Davoudi, and A.~V. Gorshkov, ``{High-Energy Collision of Quarks and
  Hadrons in the Schwinger Model: From Tensor Networks to Circuit QED},''
  \href{http://arxiv.org/abs/2307.02522}{{\ttfamily arXiv:2307.02522
  [quant-ph]}}.

\bibitem{PhysRevD.108.L091501}
K.~Ikeda, D.~E. Kharzeev, R.~Meyer, and S.~Shi, ``Detecting the critical point
  through entanglement in the schwinger model,''
  \href{http://dx.doi.org/10.1103/PhysRevD.108.L091501}{{\em Phys. Rev. D}
  {\bfseries 108} (Nov, 2023) L091501}.
  \url{https://link.aps.org/doi/10.1103/PhysRevD.108.L091501}.

\bibitem{Barata:2023jgd}
J.~a. Barata, W.~Gong, and R.~Venugopalan, ``{Realtime dynamics of hyperon spin
  correlations from string fragmentation in a deformed four-flavor Schwinger
  model},'' \href{http://arxiv.org/abs/2308.13596}{{\ttfamily arXiv:2308.13596
  [hep-ph]}}.

\bibitem{PhysRevD.108.074001}
K.~Ikeda, D.~E. Kharzeev, and S.~Shi, ``Nonlinear chiral magnetic waves,''
  \href{http://dx.doi.org/10.1103/PhysRevD.108.074001}{{\em Phys.Rev.D}
  {\bfseries 108} (Oct, 2023) 074001}.
  \url{https://link.aps.org/doi/10.1103/PhysRevD.108.074001}.

\bibitem{Lee:2023urk}
K.~Lee, J.~Mulligan, F.~Ringer, and X.~Yao, ``{Liouvillian dynamics of the open
  Schwinger model: String breaking and kinetic dissipation in a thermal
  medium},'' \href{http://dx.doi.org/10.1103/PhysRevD.108.094518}{{\em Phys.
  Rev. D} {\bfseries 108} no.~9, (2023) 094518},
  \href{http://arxiv.org/abs/2308.03878}{{\ttfamily arXiv:2308.03878
  [quant-ph]}}.

\bibitem{Ikeda:2020agk}
K.~Ikeda, D.~E. Kharzeev, and Y.~Kikuchi, ``{Real-time dynamics of Chern-Simons
  fluctuations near a critical point},''
  \href{http://dx.doi.org/10.1103/PhysRevD.103.L071502}{{\em Phys. Rev. D}
  {\bfseries 103} no.~7, (2021) L071502},
  \href{http://arxiv.org/abs/2012.02926}{{\ttfamily arXiv:2012.02926
  [hep-ph]}}.

\bibitem{itensor}
M.~Fishman, S.~R. White, and E.~M. Stoudenmire, ``{The ITensor Software Library
  for Tensor Network Calculations},''
  \href{http://dx.doi.org/10.21468/SciPostPhysCodeb.4}{{\em SciPost Phys.
  Codebases} (2022) 4}. \url{https://scipost.org/10.21468/SciPostPhysCodeb.4}.

\bibitem{itensor-r0.3}
M.~Fishman, S.~R. White, and E.~M. Stoudenmire, ``{Codebase release 0.3 for
  ITensor},'' \href{http://dx.doi.org/10.21468/SciPostPhysCodeb.4-r0.3}{{\em
  SciPost Phys. Codebases} (2022) 4--r0.3}.
  \url{https://scipost.org/10.21468/SciPostPhysCodeb.4-r0.3}.

\bibitem{PhysRevB.81.064439}
F.~Pollmann, A.~M. Turner, E.~Berg, and M.~Oshikawa, ``Entanglement spectrum of
  a topological phase in one dimension,''
  \href{http://dx.doi.org/10.1103/PhysRevB.81.064439}{{\em Phys. Rev. B}
  {\bfseries 81} (Feb, 2010) 064439}.
  \url{https://link.aps.org/doi/10.1103/PhysRevB.81.064439}.

\bibitem{Eisert:2008ur}
J.~Eisert, M.~Cramer, and M.~B. Plenio, ``{Area laws for the entanglement
  entropy - a review},''
  \href{http://dx.doi.org/10.1103/RevModPhys.82.277}{{\em Rev. Mod. Phys.}
  {\bfseries 82} (2010) 277--306},
  \href{http://arxiv.org/abs/0808.3773}{{\ttfamily arXiv:0808.3773
  [quant-ph]}}.

\bibitem{Bianchi:2021aui}
E.~Bianchi, L.~Hackl, M.~Kieburg, M.~Rigol, and L.~Vidmar, ``{Volume-Law
  Entanglement Entropy of Typical Pure Quantum States},''
  \href{http://dx.doi.org/10.1103/PRXQuantum.3.030201}{{\em PRX Quantum}
  {\bfseries 3} no.~3, (2022) 030201},
  \href{http://arxiv.org/abs/2112.06959}{{\ttfamily arXiv:2112.06959
  [quant-ph]}}.

\bibitem{Nakagawa:2017yiw}
Y.~O. Nakagawa, M.~Watanabe, S.~Sugiura, and H.~Fujita, ``{Universality in
  volume-law entanglement of scrambled pure quantum states},''
  \href{http://dx.doi.org/10.1038/s41467-018-03883-9}{{\em Nature Commun.}
  {\bfseries 9} no.~1, (2018) 1635},
  \href{http://arxiv.org/abs/1703.02993}{{\ttfamily arXiv:1703.02993
  [cond-mat.stat-mech]}}.

\bibitem{ATLAS:2023fsd}
{\bfseries ATLAS} Collaboration, G.~Aad {\em et~al.}, ``{Observation of quantum
  entanglement in top-quark pairs using the ATLAS detector},''
  \href{http://arxiv.org/abs/2311.07288}{{\ttfamily arXiv:2311.07288
  [hep-ex]}}.

\bibitem{Deutsch:2018ulr}
J.~M. Deutsch, ``{Eigenstate thermalization hypothesis},''
  \href{http://dx.doi.org/10.1088/1361-6633/aac9f1}{{\em Rept. Prog. Phys.}
  {\bfseries 81} no.~8, (2018) 082001}.

\bibitem{Kharzeev:2005iz}
D.~Kharzeev and K.~Tuchin, ``{From color glass condensate to quark gluon plasma
  through the event horizon},''
  \href{http://dx.doi.org/10.1016/j.nuclphysa.2005.03.001}{{\em Nucl. Phys. A}
  {\bfseries 753} (2005) 316--334},
  \href{http://arxiv.org/abs/hep-ph/0501234}{{\ttfamily arXiv:hep-ph/0501234}}.

\bibitem{Berges:2017hne}
J.~Berges, S.~Floerchinger, and R.~Venugopalan, ``{Dynamics of entanglement in
  expanding quantum fields},''
  \href{http://dx.doi.org/10.1007/JHEP04(2018)145}{{\em JHEP} {\bfseries 04}
  (2018) 145}, \href{http://arxiv.org/abs/1712.09362}{{\ttfamily
  arXiv:1712.09362 [hep-th]}}.

\bibitem{Zhou:2021kdl}
Z.-Y. Zhou, G.-X. Su, J.~C. Halimeh, R.~Ott, H.~Sun, P.~Hauke, B.~Yang, Z.-S.
  Yuan, J.~Berges, and J.-W. Pan, ``{Thermalization dynamics of a gauge theory
  on a quantum simulator},''
  \href{http://dx.doi.org/10.1126/science.abl6277}{{\em Science} {\bfseries
  377} no.~6603, (2022) abl6277},
  \href{http://arxiv.org/abs/2107.13563}{{\ttfamily arXiv:2107.13563
  [cond-mat.quant-gas]}}.

\bibitem{Mueller:2021gxd}
N.~Mueller, T.~V. Zache, and R.~Ott, ``{Thermalization of Gauge Theories from
  their Entanglement Spectrum},''
  \href{http://dx.doi.org/10.1103/PhysRevLett.129.011601}{{\em Phys. Rev.
  Lett.} {\bfseries 129} no.~1, (2022) 011601},
  \href{http://arxiv.org/abs/2107.11416}{{\ttfamily arXiv:2107.11416
  [quant-ph]}}.

\bibitem{Desaules:2022ibp}
J.-Y. Desaules, D.~Banerjee, A.~Hudomal, Z.~Papi\'c, A.~Sen, and J.~C. Halimeh,
  ``{Weak ergodicity breaking in the Schwinger model},''
  \href{http://dx.doi.org/10.1103/PhysRevB.107.L201105}{{\em Phys. Rev. B}
  {\bfseries 107} no.~20, (2023) L201105},
  \href{http://arxiv.org/abs/2203.08830}{{\ttfamily arXiv:2203.08830
  [cond-mat.str-el]}}.

\bibitem{Ebner:2023ixq}
L.~Ebner, B.~M\"uller, A.~Sch\"afer, C.~Seidl, and X.~Yao, ``{Eigenstate
  thermalization in (2+1)-dimensional SU(2) lattice gauge theory},''
  \href{http://dx.doi.org/10.1103/PhysRevD.109.014504}{{\em Phys. Rev. D}
  {\bfseries 109} no.~1, (2024) 014504},
  \href{http://arxiv.org/abs/2308.16202}{{\ttfamily arXiv:2308.16202
  [hep-lat]}}.

\bibitem{Yao:2023pht}
X.~Yao, ``{SU(2) gauge theory in 2+1 dimensions on a plaquette chain obeys the
  eigenstate thermalization hypothesis},''
  \href{http://dx.doi.org/10.1103/PhysRevD.108.L031504}{{\em Phys. Rev. D}
  {\bfseries 108} no.~3, (2023) L031504},
  \href{http://arxiv.org/abs/2303.14264}{{\ttfamily arXiv:2303.14264
  [hep-lat]}}.

\bibitem{Grieninger:2023ufa}
S.~Grieninger, K.~Ikeda, D.~E. Kharzeev, and I.~Zahed, ``{Entanglement in
  massive Schwinger model at finite temperature and density},''
  \href{http://dx.doi.org/10.1103/PhysRevD.109.016023}{{\em Phys. Rev. D}
  {\bfseries 109} no.~1, (2024) 016023},
  \href{http://arxiv.org/abs/2312.03172}{{\ttfamily arXiv:2312.03172
  [hep-th]}}.

\bibitem{Florio:2023mzk}
A.~Florio, ``{Two-fermion negativity and confinement in the Schwinger model},''
  \href{http://arxiv.org/abs/2312.05298}{{\ttfamily arXiv:2312.05298
  [hep-th]}}.

\end{thebibliography}\endgroup

\end{document}